\documentclass[iicol, sn-mathphys-num]{sn-jnl}%
\usepackage{amsfonts,amssymb}
\usepackage{array}
\usepackage{pgfplots}
\usepackage{verbatim}
\usetikzlibrary{decorations.markings}
\usepackage{enumerate}
\usepackage{colortbl}
\usepackage{booktabs}\let\cline\cmidrule
\usepgfplotslibrary{units}
\usetikzlibrary{spy,backgrounds}
\usetikzlibrary{shapes.geometric} 
\usepackage{pgfplotstable}
\usepackage{flushend}
\usepackage[normalem]{ulem}
\usepackage{multirow}
\usepackage{nicefrac} 
\usepackage[acronym]{glossaries}

\hypersetup{%
  colorlinks=true,  
  linkcolor=black, 
  linktoc=all,
}

\newcommand{\PM}{\text{PM}}
\newcommand{\FM}{\text{FM}}
\newcommand{\Ical}{\mathcal{I}}
\newcommand{\Icalsort}{\mathcal{I}_{\text{sort}}}
\newcommand{\Lij}[2]{\lambda_{#1,#2}}

\newcommand{\Fcal}{\mathcal{F}}

\newcommand{\latSC}{\mathcal{L}_{\text{SC}}}
\newcommand{\latSCF}{\mathcal{L}_{\text{SCF}}}

\newcommand{\avgSCF}{\overline{\mathcal{L}_{\text{SCF}}}}
\newcommand{\avgSCLF}{\overline{\mathcal{L}_{\text{SCLF}}}}

\newcommand{\Tmax}{T_{\text{max}}}
\newcommand{\prob}[2]{\mathbb{P}_{#2}\left(#1\right)}
\newcommand{\cbf}{\mathbf{c}}
\newcommand{\ebf}{\mathbf{e}}

\newcommand{\latSCL}{\mathcal{L}_{\text{SCL}}}
\newcommand{\latPSCLF}{\mathcal{L}_{\text{PSCLF}}}
\newcommand{\avgPSCLF}{\overline{\mathcal{L}_{\text{PSCLF}}}}

\newcommand{\avgt}{\overline{t}}

\definecolor{matlab1}{rgb}{0.000, 0.447, 0.741} 
\definecolor{matlab2}{rgb}{0.850, 0.325, 0.098} 
\definecolor{matlab3}{rgb}{0.929, 0.694, 0.125} 
\definecolor{matlab4}{rgb}{0.494, 0.184, 0.556} 
\definecolor{matlab5}{rgb}{0.466, 0.674, 0.188} 
\definecolor{matlab6}{rgb}{0.301, 0.745, 0.933} 
\definecolor{matlab7}{rgb}{0.635, 0.078, 0.184} 

\newacronym{awgn}{AWGN}{additive white Gaussian noise}
\newacronym{bpsk}{BPSK}{Binary Phase-Shift Keying}
\newacronym{snr}{SNR}{signal-to-noise ratio}
\newacronym{fer}{FER}{frame-error rate}
\newacronym{sc}{SC}{Successive-Cancellation}
\newacronym{scl}{SCL}{SC List}
\newacronym{scf}{SCF}{SC Flip}
\newacronym{sclf}{SCLF}{Successive-Cancellation List Flip}
\newacronym{dscf}{DSCF}{Dynamic SCF}
\newacronym{pscf}{PSCF}{Partitioned SCF}
\newacronym{cdf}{CDF}{Cumulative Density Function}
\newacronym{pscl}{PSCL}{Partitioned SCL}
\newacronym{psclf}{PSCLF}{Partitioned SCLF}
\newacronym{crc}{CRC}{cyclic-redundancy check}
\newacronym{ca}{CA}{CRC-aided}
\newacronym{llr}{LLR}{log-likelihood ratio}
\newacronym{cc}{CC}{clock cycle}
\newacronym{srm}{SRM}{simplified restart mechanism}
\newacronym{grm}{GRM}{generalized restart mechanism}
\newacronym{ascl}{ASCL}{Adaptive-SCL}
\newacronym{cr}{CR}{Check and Remove}
\newacronym{ck}{CK}{Check and Keep}
\newacronym{llrm}{LLRM}{Limited Location Restart Mechanism}
\begin{document}

\title{On Reducing Decoding Complexity of Successive-Cancellation List Flip Decoding of Polar Codes}
	
\author*[1]{\fnm{Charles} \sur{Pillet}}\email{charles.pillet@lacime.etsmtl.ca}
\author*[1]{\fnm{Ilshat} \sur{Sagitov}}\email{ilshat.sagitov.1@ens.etsmtl.ca}
\author*[1]{\fnm{Pascal} \sur{Giard}}\email{pascal.giard@etsmtl.ca}
\affil*[1]{\orgdiv{LaCIME, Department of Electrical Engineering}, \orgname{\'Ecole de technologie sup\'erieure (\'ETS}, \orgaddress{\street{1100 Notre-Dame St West}, \city{Montr\'eal}, \postcode{H3C 1K3}, \state{Qu\'ebec}, \country{Canada}}}

\abstract{
    The recently proposed \gls{sclf} decoding algorithm for polar codes improves the error-correcting performance of state-of-the-art \gls{scl} decoding. 
    However, it comes at the cost of a higher complexity.
    In this paper, partitioned polar codes tailored for the proposed \gls{psclf} decoding algorithm are used to reduce the complexity of \gls{sclf}.
    Indeed, compared to \gls{sclf}, \gls{psclf} allows early termination and is able to restart by skipping part of the decoding tree traversed sequentially.  
    In order to maximize the coding gain,  design of partitions tailored to \gls{psclf} is proposed. 
    In this extended paper, dynamic flip metric is used, as well as the possibility to flip multiple times during \gls*{scl}.
    An analysis on the impact of this strategy on the early-termination or the \gls{crc} collisions encountered in \gls{psclf} is carried out.
    Error-correction performance of multiple code rates and multiple partition strategies are shown.
    With the baseline algorithm SCL with $L=2$, degradation of $0.05$ dB is shown with respect to \gls{scl}-64, using $\omega=3$ flip per trial with $\Tmax=300$ trials.
    Numerical results show that the proposed \gls{psclf} algorithm has an error-correction performance gain of up to 0.1\,dB with respect to \gls{sclf} with same decoding parameters.
    This work is also compared with existing techniques to reduce the complexity of the \gls*{sclf} decoding algorithm.
    The proposed algorithm reduces the complexity up to $77$\,\% at the frame-error rate of $0.01$ with respect to \gls{sclf} and is able to reduce more the decoding complexity of \gls{sclf} embedding as well a restart mechanism.
    The average execution time of \gls{psclf} matches the latency of \gls{scl} at $\text{FER}=\mathbf{4\cdot10^{-3}}$ and lower.    
   }

\maketitle
\glsresetall

\section{Introduction}
    Since the joint invention of polar codes and the asymptotically capacity-achieving \gls{sc} decoding algorithm \cite{ArikanPolarCodes}, progress towards improving the error-correcting performance of \gls{sc} at finite block length has been made by proposing new decoding algorithms of polar codes \cite{SCL,scf_intro}.
    \gls{scl} is the list decoding algorithm based on \gls{sc} \cite{SCL}.
    \gls{scl} tracks in parallel a list of $L$ candidates which improves the error-correcting performance.
    The error-correcting performance of \gls{scl} can further be improved by a concatenation of a \gls{crc} code. 
    This scheme is referred as \gls{ca}-polar codes and has been chosen as one of the coding scheme in the 5G standard \cite{standard}.
    
    An alternative decoding algorithm of \gls{ca}-polar codes based on \gls{sc} decoding is \gls{scf} \cite{scf_intro}.
    While the list of candidates was generated in parallel for \gls{scl}, the list of candidates is generated sequentially for \gls{scf} by flipping potential error-prone bits after the first \gls{sc} trial.
    The accuracy of identifying error-prone bits improved in \cite{dyn_scf} while also enabling multi-flipping in additional \gls{sc} trials. 
    \gls{scf} and its variants have a variable execution time, but a throughput and a complexity asymptotically equal to those of \gls{sc}.

    \gls{sclf} is the flip decoding algorithm based on \gls{scl} \cite{first_SCLF}.
    If the first \gls{scl} fails, a list of path-flipping locations is retrieved and \gls{scl} is performed once more and take the $L$ worst paths at the flip location \cite{flip_criteria_real_time,flip_criteria_fixed}.
    Various flip metrics have been proposed to find the path-flipping locations.
    A reliable flipping set based on a heuristic parameter is proposed \cite{metric_with_alpha}.
    Authors in \cite{reduced_complexity_metric} reduce drastically the complexity of this flip metric.
    A dynamic flipping set permitting to adjust the candidate flip locations was proposed in \cite{dyn_sclf}.
    \gls{sclf} can support multiple flip locations \cite{metric_with_alpha,dyn_sclf} which permits to improve even the error-correction performance at the cost of increasing the decoding complexity.
    By combining list and flip decoding strategy, \gls{sclf} returns the state-of-the-art error-correcting performance at the cost of increased complexity and variable execution time.

    Partitioned polar codes \cite{partition_scl,segmented_scl} are a special type of \gls{ca}-polar codes.
    Partitioned polar codes are segmented into partitions, each of which is protected by its own \gls{crc}.
    \gls{scl} \cite{partition_scl,segmented_scl} and \gls{scf} \cite{PSCF} support the decoding of partitioned polar codes and are referred to as \gls{pscl} and \gls{pscf}.
    For \gls{pscl}, a coding gain with respect to \gls{scl} has been observed \cite{partition_scl,segmented_scl}.
    Moreover, at equal error-correcting performance, the decoding complexity is reduced \cite{segmented_scl} as well as the memory requirements if the partitions correspond to sub-decoding trees \cite{partition_scl}.
    For \gls{pscf}, the number of flipping trials has been shown to be dividable by a factor $4$ with respect to \gls{scf} for an equivalent error-correcting performance \cite{PSCF}.
    In \cite{PPC}, partitioned polar codes are decoded with the simplest \gls{sclf} decoding algorithm, i.e., the dynamic flipping strategy is not implemented.
    In order to maximise the coding gain with respect to \gls{sclf}, the partitions are designed according to the main decoder being SCL.
    This approach improves error-correcting performance with respect to other naive partition design strategies \cite{partition_scl,segmented_scl}.
    \gls{psclf} permits to reduce the decoding complexity of \gls{sclf} while improving the decoding performance.

    This paper is an extension of \cite{PPC}.
    In this paper, the dynamic flipping strategy is investigated for the \gls{psclf} decoding algorithm.
    We analyse the impact of this flipping strategy on the decoding behavior of \gls{psclf}.
    Namely, the probability of \gls{crc} collisions and of the early-termination are provided for all possible decoding parameters.
    It highlights that the proposed decoding strategy is likely to be subject to \gls{crc} collisions, especially with the newly proposed flipping strategy.    
    Analysis of the decoding with different \gls{crc} structures is performed to conclude that the simplest \gls{crc} structure is a decent choice.
    With respect to \cite{PPC}, we propose another restart strategy when the decoding moves to the next partition, improving error-correcting performance at no cost.
    Finally with respect to \cite{PPC}, the average execution time of \gls{psclf}  is compared to \gls{sclf} embedding techniques reducing its average execution time. 
    Analysis is performed for all flipping and partition strategies.    

\section{Preliminaries}

\subsection{Polar Codes}
    A $(N=2^n,K)$ polar code of length $N$ and dimension $K$ is a binary block code based on the polarization effect of the binary kernel $\mathbf{T}_2=\left[\begin{smallmatrix}
        1&0\\1&1
    \end{smallmatrix}\right]$ and of the transformation matrix $\mathbf{T}_N=\mathbf{T}_2^{\otimes n}\in\mathbb{F}_2^{N\times N}$ \cite{ArikanPolarCodes}.
    A $(N,K)$ polar code is fully defined by its information set $\Ical\subseteq[N]\triangleq\{0,\dots,N-1\}$, describing the locations where the message $\mathbf{m}\in\mathbb{F}_2^K$ is inserted in the input vector $\mathbf{u}=(u_0,\dots,u_{N-1})\in\mathbb{F}_2^N$, i.e., $\mathbf{u}_\Ical=\mathbf{m}$.
    The remaining $N-K$ locations, stored in the frozen set $\Fcal=[N]\setminus\Ical$ are set to 0, i.e., $\mathbf{u}_\Fcal=\mathbf{0}$.
    The encoding is performed as $\mathbf{x}=\mathbf{u}\cdot\mathbf{G}_N$, where $\mathbf{x}\in\mathbb{F}_2^N$ is a codeword.

    A $(N,K+C)$ \gls{ca}-polar code is a  $(N,K)$ polar code concatenated with a \gls{crc} code of $C$ bits.
    The \gls{crc} encoder is applied on $\mathbf{m}$ and $C$ \gls{crc} bits are appended to $\mathbf{m}$, defining $\mathbf{m}'\in\mathbb{F}_2^{K+C}$.
    Hence, $\Ical$ is now enlarged by $C$ additional bits.
    As a rule, we have $\Ical=\{i_1,i_2,\dots,i_{K+C}\}$ with $i_1<i_2<\dots<i_{K+C}$.
    In the remaining of the paper, \gls{ca}-polar codes are used and decoded with \gls{sc}-based  \cite{ArikanPolarCodes} algorithms.
    \gls{sc} leads to poor error-correcting performance in the finite-length regime but can be improved with more complex \gls{sc}-based decoding algorithm.

	\subsection{SCL Decoding}
    \gls{scl} is the list decoding algorithm also based on \gls{sc}  \cite{SCL}.
    Hence the scheduling is as \gls{sc} and at each information bit $i\in\Ical$, the paths consider both possible values $\{0,1\}$, doubling the size of the list.
    In order to limit the complexity, \gls{scl} only considers $L$ different decoding paths such that path sorting according to a metric is required.
    If $i\in\{i_1,\dots,i_{\log_2(L)}\}$, no path sorting is needed since doubling the number of candidates still generates less than or exactly $L$ candidate paths.
    However, if $i\in\{i_{\log_2(L)+1}, \dots,i_{K+C}\} \triangleq \Icalsort$, the $L$ paths minimizing the path metrics out of the $2L$ candidate paths are selected \cite{SCL_LLR}.
    The path metric $\PM_i[l]$ for index $i\in[N]$ and path $l\in[2L]$ is penalized whenever the bit decision $\hat{u}_i[l]$ of the partial candidate vector $\hat{\mathbf{u}}_0^i[l]\in\mathbb{F}_2^{i+1}$ does not correspond to the hard decision of the \gls{llr} $\Lij{0}{i}[l]$ in the $i^{th}$ leaf of path $l$.
    \gls{scl} returns $L$ candidates for $\mathbf{u}$, noted $\hat{\mathbf{u}}[l]$ with $0\leq l\leq L-1$.
    The candidate passing the \gls{crc} with the lowest path metric is chosen as decoder output $\mathbf{\hat{u}}$ which improves the error-correction performance.

    \gls{ascl} \cite{ASCL} is an iterative decoding algorithm successively increasing the list size $L$ up to a maximum and stops as soon as the CRC code is checked.
    The complexity of \gls{ascl} converges to \gls{sc} at low \gls{fer}, but the time complexity increases and depends from the channel condition.
    \subsection{SCF Decoding}
    \gls{scf} decoding performs up to $T_{\max}$ SC decoding trials. 
    A \gls{crc} check is performed at the end of each trial. 
    If the first trial fails, a set of flipping candidates $\beta$ is generated with $|\beta|=T_{\max}-1$. 
    The $t^{th}$ additional trial performs SC except at the flipping location $\beta_t$ where the reverse decision on $\hat{u}_{\beta_{t}}$ is taken.
    The decoding latency of \gls{scf} is $\latSCF =T_{\max}\times \latSC$ while its average execution time $\avgSCF$ corresponds to
    \begin{align}
        \avgSCF = \avgt \times \latSC \leq \latSCF,\label{eq:avgSCF}
    \end{align}
    where $\avgt\leq T_{\max}$ is the average number of decoding trials.
    
    \gls{pscf} \cite{PSCF} is a flip decoder allowing to decode \gls{ca}-polar codes with multiple \glspl{crc} distributed in the codeword.
    A partitioned polar code is partly described by its  set $\mu$ corresponding to the last indices of all partitions.
    The size $P=|\mu|$ describes the number of partitions.

    \subsection{SCLF Decoding}
    \gls{sclf} algorithm \cite{first_SCLF} is the flip decoding algorithm of polar codes with \gls{scl} as the core decoder. 
    The best flipping strategy has been proposed in \cite{flip_criteria_fixed,flip_criteria_real_time}, i.e., on a flip location $i\in\Icalsort$, the $L$ worst paths are selected instead of the best $L$ paths \cite{SCL}.

    If we note the flip metric $\FM_i^\alpha$ at index $i\in\Icalsort$ and $\alpha\geq 1$, a factor mitigating the impact of propagation of errors. 
    A reliable flip metric was proposed in \cite{metric_with_alpha}, later simplified for hardware purposes in \cite{reduced_complexity_metric} as
    \begin{align}\label{eq:reduced_complexity_metric}
        \FM_i^\alpha=-\PM_i[0]+\alpha\PM_i[L],
    \end{align}
    where the path metrics are sorted from most to least reliable.
    Since a lower complexity \gls{sclf} variant is targeted, \eqref{eq:reduced_complexity_metric} is selected.
    The flipping set $\mathcal{B}_{\text{flip}}\subset\Icalsort$, storing the indices where the flips are performed, is designed such that $\mathcal{B}_{\text{flip}}(t)$ is the index $i\in\Icalsort$ with the $t^{th}$ lowest flip metric \cite{metric_with_alpha}.
    This flipping strategy were used in \cite{PPC}.
    In this paper, the dynamic flipping strategy is used as in \cite{LLRM}.
    Namely, we use \eqref{eq:reduced_complexity_metric} as based metric with $\alpha=1$ in order to generate the flip metric of a set of flipping indices of size lower or equal to $\omega$, being the decoding order of \gls{psclf}, i.e., the maximum number of path flips in an additional trial.
    
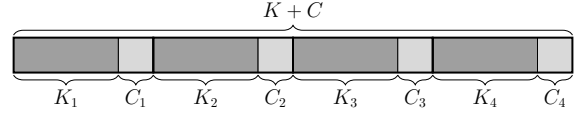
\begin{figure}[t]
    \centering
    \resizebox{.99\columnwidth}{!}{\usetikzlibrary{decorations.pathreplacing}

\begin{tikzpicture}
    \newcommand{\height}{1}
    \draw (0,0) rectangle (16,\height);
    
    \draw[fill=gray!75] (0,0) rectangle (3,\height);
    \draw[fill=gray!30] (3,0) rectangle (4,\height);
    
    \draw[fill=gray!75] (4,0) rectangle (7,\height);
    \draw[fill=gray!30] (7,0) rectangle (8,\height);
    \draw[fill=gray!75] (8,0) rectangle (11,\height);
    \draw[fill=gray!30] (11,0) rectangle (12,\height);
    \draw[fill=gray!75] (12,0) rectangle (15,\height);
    \draw[fill=gray!30] (15,0) rectangle (16,\height);

    \draw[ultra thick] (0,0) rectangle (4,\height);
    \draw[ultra thick] (4,0) rectangle (8,\height);
    \draw[ultra thick] (8,0) rectangle (12,\height);
    \draw[ultra thick] (12,0) rectangle (16,\height);
    \draw [decorate,decoration={brace,amplitude=8pt,mirror,raise=2pt}] (0,0) -- (3,0) node[midway,below=10pt] {\Large$ K_1$};
    \draw [decorate,decoration={brace,amplitude=8pt,mirror,raise=2pt}] (3,0) -- (4,0) node[midway,below=10pt] {\Large$ C_1$};
    \draw [decorate,decoration={brace,amplitude=8pt,mirror,raise=2pt}] (4,0) -- (7,0) node[midway,below=10pt] {\Large$ K_2$};
    \draw [decorate,decoration={brace,amplitude=8pt,mirror,raise=2pt}] (7,0) -- (8,0) node[midway,below=10pt] {\Large$ C_2$};
    \draw [decorate,decoration={brace,amplitude=8pt,mirror,raise=2pt}] (8,0) -- (11,0) node[midway,below=10pt] {\Large$ K_3$};
    \draw [decorate,decoration={brace,amplitude=8pt,mirror,raise=2pt}] (11,0) -- (12,0) node[midway,below=10pt] {\Large$ C_3$};
    \draw [decorate,decoration={brace,amplitude=8pt,mirror,raise=2pt}] (12,0) -- (15,0) node[midway,below=10pt] {\Large$ K_4$};
    \draw [decorate,decoration={brace,amplitude=8pt,mirror,raise=2pt}] (15,0) -- (16,0) node[midway,below=10pt] {\Large$ C_4$};
    \draw [decorate,decoration={brace,amplitude=8pt,raise=2pt}] (0,\height) -- (16,\height) node[midway,above=12pt] {\Large$ K+C$};
    
\end{tikzpicture}}
    \caption{Message $\mathbf{m}'\in\mathbb{F}_2^{K+C}$ allocated in $\Ical$ for a partitioned polar code with $P=4$ partitions. Dark gray corresponds to the message $\mathbf{m}$ and light gray corresponds to the \gls{crc} bits.}
    \label{fig:message_partitione}
\end{figure}

\section{Partitioned SCLF decoder}
Both \gls{scl} and \gls{scf} have been adapted to decode partitioned polar codes \cite{partition_scl,segmented_scl,PSCF}. 
The \gls{pscl} and \gls{pscf} algorithms  were shown to have an improved error-correction performance and a reduced decoding complexity over their respective counterparts.
\gls{sclf} is a more complex algorithm than \gls{scf} or \gls{scl}, hence we propose the \gls{psclf} algorithm and study the complexity reduction.
\subsection{Partitioned Polar Codes}
A partitioned polar code is divided into $P>1$ partitions. 
The $p^{th}$ should contain $K_p\geq 1$ information bits and is concatenated with its own \gls{crc} code of size $C_p$.
The structure of \gls{crc} codes is noted $\mathcal{C}=\{C_1,\dots,C_P\}$.
The number of information bits and \gls{crc} bits verify $K=\sum_{p=1}^P K_p$ and $C=\sum_{p=1}^P C_p$.
Thus, the partitioned polar code is as well defined with $|\Fcal|=N-K-C$ and $|\Ical|=K+C$.
\autoref{fig:message_partitione} depicts the message $\mathbf{m}'\in\mathbb{F}_2^{K+C}$ for $P=4$ partitions.
The vector $\mathbf{m}'$ is then allocated in the positions stated in $\Ical$, retrieving $\mathbf{u}$ and the polar encoding is performed to retrieve the codeword $\mathbf{x}$.

In the following, the number of non-frozen bits in the $p^{\text{th}}$ partition is noted $s_p=K_p+C_p$ while the cumulative number of non-frozen bits is noted $S_p=\sum_{i=1}^{p}s_p$, hence we can express the partition design as 
\begin{align}
    \mathcal{K}=\{S_1=s_1,\dots,S_p=K+C\}\label{eq:KPval}
\end{align}
Similarly, the partition design can be expressed with the last indices of the partitions inside the codeword, namely, 
\begin{align}
    \mu=\{\mu_1,\dots, \mu_P\},
\end{align} 
with $\mu_p\triangleq\Ical\left(S_p\right)=i_{S_p}$, $1\leq p\leq P$.
Given that $N-1$ belongs to $\Ical$, we have 
\begin{align}
    \mu_P=N-1\label{eq:muPval}
\end{align}

An elementary design of partitions $\mu$ is to uniformly distribute the partitions based on the information set $\Ical$ \cite{segmented_scl}.
Namely, it corresponds to 
\begin{align}
    \mathcal{K}_{\text{divK}}=\left\{\frac{K+C}{P},\frac{2\cdot\left(K+C\right)}{P},\dots,K+C\right\}.\label{eq:defdivK}
\end{align}
For a code $(1024,512+32)$ and $P=4$, each partition contains $\frac{K+C}{P}=136$ non-frozen bits.
Another simple design of the partition $\mu$ is to consider partitions as sub-decoding trees \cite{partition_scl}. 
This design corresponds to
\begin{align}
    \mu_{\text{divN}}=\left\{\frac{N}{P}-1,\frac{2\cdot\left(N\right)}{P}-1,\dots,N-1\right\}.\label{eq:defdivN}
\end{align}
This approach is particularly valid for $P=2^x$ as it will divide equally the decoding tree in $P$ sub-decoding tree of same size and structure. 
However, this approach may lead to partitions without information bits for low-rate codes. This approach was not considered in \cite{PPC}.

\subsection{Description of \gls{psclf} Decoding}\label{subsec:psclf_description_crck}
\Gls{psclf} decoding requires knowledge of the \gls{crc} structure $\mathcal{C}$ and the partition $\mu$.
On each partition, up to $T_{\max}$ \gls{scl} trials are performed.
In \cite{PPC}, the simplest flipping strategy was used, i.e., the flip metric is not dynamically updated.
In this extended paper, the flip metric is now dynamically updated and up to $\omega=3$ path flips can be performed, similarly to the dynamic flip method described in \cite{LLRM}.
At a flipping position, the reverse decision on the paths is taken, i.e., the decoding pursues with the worst $L$ paths instead of the best $L$ ones.
After reaching $\mu_1$ for the first time, the flipping set $\beta$ is computed only when none of the paths satisfy the \gls{crc} code.
If $T_{\max}$ trials are performed and still no paths satisfy the \gls{crc}, decoding failure is raised.
If at a trial $1\leq t \leq T_{\max}$, the \gls{crc} is checked for at least one path, the decoding of the next partition begins.

When reaching the end of the partition and having a valid \gls{crc}, the \gls{ck} method described in \cite{5G_DCA_WCNC} was used in the original \gls{psclf} paper \cite{PPC}.
This method provides the most simple approach to pursue the decoding when \gls{crc} codes are distributed in the codeword. 
Namely, both paths that validate the \gls{crc} and those that do not are kept and forwarded to the next partition.
This method corresponds to the original \gls{scl} algorithm since no changes are made to the algorithm.
However,  it suffers from a small performance degradation compared to other methods requiring more data control \cite{5G_DCA_WCNC}.

In this extended paper, the \gls{cr} method \cite{5G_DCA_WCNC} is also investigated.
During the restart of a partition, only the paths verifying the \gls{crc} at the previous partition remain. 
Hence, all decoding paths not verifying the \gls{crc} are discarded at the beginning of the partition, leading to additional implementation complexity.
In order to provide a simple implementation of this method, at the end of a partition, if at least one path validates \gls{crc}, the $L$ path metrics are updated as 
\begin{align}
    \PM\left(l\right) = 
    \begin{cases}
    \PM\left(l\right) + O,& \text{if}\,\, c_c(l)=0 \\
    \PM\left(l\right), \,& \text{otherwise},
    \label{eq:pm_scl_CR}
    \end{cases}
\end{align}
where $O$ is a large value constant and $c_c(l)=0$ states that the $l^{th}$ path did not pass the \gls{crc}.
This simple implementation discards the wrongly estimated paths by artificially penalizing their path metrics.
Indeed, a small path metric means that the path is reliable \cite{SCL_LLR}.
At the next information bit of the new partition, the surviving paths will be the ones duplicated through the path verifying the \gls{crc} since its path metric was not penalized and kept to its original value \eqref{eq:pm_scl_CR}.

Decoding is successful if each partition has a path passing the \gls{crc}. 
\Gls{psclf} naturally implements \emph{early termination} as decoding stops before the end of a frame whenever a decoding failure happens at a partition $1\leq p<P$.
Early-termination is described in more details in \autoref{subsec:et_psclf}.

\subsection{Average Execution Time of \gls{psclf}}\label{subsec:exec_psclf}
In \cite{PSCF}, the complexity of \gls{pscf} was linked to the \emph{normalized average computational complexity}. 
However, no equations were given to compute it.
In the following, we provide detailed equations to compute the average execution time of \gls{psclf}.
The average execution time of \gls{psclf} refers to the average amount of time that \gls{psclf} takes to complete its execution.
It is computed by way of simulations.
Since the partitions are of different lengths, the impact of flipping fluctuates from partition to partition.
The partial latency $\latSC\left(i\right)$ required by the semi-parallel SC decoder to decode until the bit $i\in[N]$ is derived from \cite{GRM_TSP}
\begin{align}
 \label{eq:l_p_saved_calc}
 \latSC\left(i\right) &= \sum_{s=0}^{n-1}\left\lceil \frac{2^s}{\varphi}\right  \rceil +\sum_{s=0}^{n-1} \left( \left \lceil \frac{2^s}{\varphi}\right  \rceil \times \left \lfloor \frac{i}{2^s} \right \rfloor \right),
\end{align}
where $\varphi$ is the number of processing elements.
If $i=N-1$ and corresponds to the end of the decoding, the latency of \gls{sc} is retrieved, i.e., $\latSC=\left(2N + \frac{N}{\varphi}\cdot \log_2{\left(\frac{N}{4\varphi}\right)}\right)$ \cite{semi_par_sc}.

In this paper, we consider that the SCL implementation is fully parallel, based on this assumption, the latency of \gls{scl} is independent of the list size $L$ and is given in \cite{PolarBear} as 
\begin{align}
    \latSCL=\latSCL(N-1)=\latSC+|\Ical|,\label{eq:latSCL_all}
\end{align} in which the last term corresponds to the latency incurred by the sorting steps.
The partial \gls{scl} latency $\latSCL\left(i\right)$ depends on the number of sorting operations required up to index $i$. 
Given that $\mathcal{K}=\{s_1,\dots,s_P\}$ is the set storing the number of non-frozen bits in each partition, the \gls{scl} partial latency $\latSCL\left(\mu_p\right)$ to decode the first $p$ partitions is
\begin{align}
 \label{eq:latSCL}
 \latSCL\left(\mu_p\right) &= \sum_{m=1}^{p}s_m + \latSC(\mu_p),
\end{align}
where the left term represents the number of sorting operations by the end of the $p^{\text{th}}$ partition.

Similarly to \gls{scf}'s \eqref{eq:avgSCF}, the average execution time of \gls{sclf} is $\avgSCLF=\latSCL\times\overline{t}$ where $\overline{t}$ is the average number of \gls{scl} trials per frame.
However, to compute the average execution time of \gls{psclf}, the early-termination property needs to be taken into account.
Next, the probability that the decoding is performed in partition $p$ is noted $\mathbb{P}(T_p)$. 
The probability is linked to the early-termination since it corresponds to the probability that decoding has not stopped early in any of the previous $p-1$ partitions.
For the first partition, the probability is $\mathbb{P}(T_1)=1$.
The average execution time of \gls{psclf} corresponds to the summation of the average time spent in each partition mitigated by the probability $\mathbb{P}(T_p)$ of decoding it:
\begin{align}
    \avgPSCLF = \sum_{p=1}^{P} \mathbb{P}(T_p)\,\overline{t_p}\left(\latSCL(\mu_p)-\latSCL(\mu_{p-1})\right)\label{eq:avgPSCLF},
\end{align}
where $1\leq\overline{t_p}\leq T_{\max}$ is the average number of \gls{scl} trials to decode the $p^{\text{th}}$ partition and by convention $\latSCL(\mu_{0}=0)=0$. 

The latency of \gls{psclf}, corresponding to the worst execution time possible, is achieved when the decoding of each partition terminates at the last trial $\Tmax$, i.e., the CRC code is checked at the last trial for all $P$ partitions.
Hence, the latency of \gls{psclf} $\latPSCLF$ is
\begin{align}
    \latPSCLF&=\sum_{p=1}^{P} \Tmax\left(\latSCL(\mu_p)-\latSCL(\mu_{p-1})\right)\label{eq:latPSCLF},\\
    \latPSCLF&=\Tmax\cdot\latSCL.
\end{align}
In opposite to flip decoders \cite{dyn_scf,dyn_sclf}, the proposed \gls{psclf} decoding algorithm is unlikely to return an average execution time close to its latency when the channel condition is poor.
In fact, the proposed \gls{psclf} algorithm will either early-terminate, i.e., the algorithm fails to check the \gls{crc} code in one of the first partitions, or it will encounter \gls{crc} collisions prior the $\Tmax^{th}$ trial as discussed in the next section.
Both reduce the execution time of \gls{psclf} with respect to the latency in poor channel condition.
Hence, the average execution time of \gls{psclf} will be lower than its latency. It will be verified in \autoref{subsec:avg_exec_time_sim}.
\subsection{Memory Model of PSCLF}
In \cite{PPC}, the memory model of PSCLF was omitted due to space limitation.
The proposed \gls{psclf} shares similarities with the work presented in \cite{LLRM}, namely, both algorithms are \gls{sclf} with multiple path flip per trials, and the restart is not performed at index 0.
In \cite{LLRM}, the restart during a flip trial is performed in one of the restart locations defined prior the decoding.
In \gls{psclf}, the restart is performed in the first index of each partition in \gls{psclf}. also being known prior the decoding. 
For the first partition, the restart can be performed in the first index in $\Ical$ as described in \cite{LLRM}. 

In order to restart at the start of each partition when all $L$ \gls{crc} fail, information is required to be stored at the beginning of the partition.
The information corresponds to the $2L$ partial messages candidates, storing the message candidates for all $2L$ paths, as well as the $2L$ path metrics associated to the $2L$ paths.
The $2L$ path information and not only the $L$ best path information is required, indeed, if a path flipping is required at the restart location, it will be impossible to restart since  the information of the $L$ worst paths is needed.
Finally, the position of the start of each partition is required.
The check and remove variant \eqref{eq:pm_scl_CR} intervenes as well in the start of the partition, however, the memory overhead caused by the check and remove is negligible.

In \cite{LLRM}, an estimation of the memory overhead was performed, since \gls{psclf} shares the same structure of memory overhead, the overhead is similar.
Namely, the overhead is not greater than $15.0\%$ and can be as small as $1.1\%$.

\section{On CRC collisions in PSCLF}\label{sec:collisions}
In \cite{PPC},  partitioned polar codes decoded by \gls{psclf} are shown to encounter many \gls{crc} collisions, hence, a \gls{crc} structure based on the decoding capability of the scheme was proposed.
It was shown that the probability of errors shifts towards the end when the channel condition improves, as depicted in \autoref{fig:CDF_SNR}, hence, a longer \gls{crc} code was attributed to the last partition.
However, no probability analysis of the collisions was provided.

Next, a probability analysis of \gls{crc} collisions under low SNR condition is provided.
Based on the probability analysis, a study of the early-termination in \gls{psclf} is also given.
\subsection{CRC Collisions Under Low SNR Condition}
In poor channel condition, no decoding algorithm will return a good candidate, in this scenario, we assume that the decoding algorithm returns random bit sequences as decoding candidates.
Next, given a bit sequence, a \gls{crc} collision corresponds to the event where the bit sequence passes the \gls{crc}, despite not being the correct sequence.
The event of having at least one CRC collision is noted $\cbf$ in the following. 
Since the candidate is considered to be randomly generated, the second condition, i.e., the sequence is not correct, is met if a \gls{crc} collision occurs since the candidate is one of the $2^K$ possible candidates.

Next, the generated \gls{crc} is considered to be of equal probability.
Namely, given a randomly generated bit sequence protected by a \gls{crc} code of length $C$ will induce a collision with probability $\prob{\cbf}{C}$
\begin{align}
    \prob{\cbf}{C}=\frac{1}{2^{C}}\label{eq:crc_collision}
\end{align}
since there are $2^{C}$ possible check values.
Similarly, the probability $\prob{\overline{\cbf}}{C}$ to fail the \gls{crc}, i.e., not facing a \gls{crc} collision, is 
\begin{align}
    \prob{\overline{\cbf}}{C} &= 1-\prob{\cbf}{C}=\frac{2^C-1}{2^C}.
\end{align}
Given \eqref{eq:crc_collision}, the chance of \gls{crc} collisions increases with smaller \gls{crc} codes, i.e., given two \gls{crc} codes of size $C_1<C_2$, we have
\begin{align}
    \prob{\cbf}{C_1}=\frac{1}{2^{C_1}}>\frac{1}{2^{C_2}}=\prob{\cbf}{C_2}.\label{eq:probC1C2}
\end{align}
For instance, for $C_1=1$, the probability is of $\prob{\cbf}{C_1}=\nicefrac{1}{2}$, while for $C_2=8$, $\prob{\cbf}{C_2}=\nicefrac{1}{256}$. 

If not only one but $L>1\in\mathbb{N}$ bit sequences are randomly generated, the probability $\prob{\cbf}{C,L}$ to face up to at least one \gls{crc} collision is
\begin{align}
    \prob{\cbf}{C,L}&=1-\prob{\overline{\cbf}}{C,L}\\
    \prob{\cbf}{C,L}&=1-\left(\prob{\overline{\cbf}}{C}\right)^{L},
\end{align}
where $\prob{\overline{\cbf}}{C,L}$ is the probability that all $L$ bit sequences fail the \gls{crc} check.
We note that
\begin{align}
    \prob{\cbf}{C} < \prob{\cbf}{C,L},\label{eq:CRCcollision_higher_chance_L}
\end{align}
because $\prob{\overline{\cbf}}{C,L}<\prob{\overline{\cbf}}{C}$ since $\prob{\overline{\cbf}}{C}<1$.
Similarly, given a number of trials $\Tmax$, the probability $\prob{\cbf}{C,L,\Tmax}$ that at least one \gls{crc} collision happens during one of the $\Tmax$ successive trials is 
\begin{align}
    \prob{\cbf}{C,L,\Tmax}&=1-\prob{\overline{\cbf}}{C,L,\Tmax},\label{eq:atleastonecol}\\
    \prob{\cbf}{C,L,\Tmax} &=1-\left(\prob{\overline{\cbf}}{C}\right)^{L\times\Tmax} 
\end{align}
where $\prob{\overline{\cbf}}{C,L,\Tmax}$ is the probability that no collision happens for $\Tmax$ successive trials with $L$ sequences generated per trial.
As for \eqref{eq:CRCcollision_higher_chance_L}, increasing the number of successive trials also increase the probability of \gls{crc} collisions, namely,
\begin{align}
    \prob{\cbf}{C,L} < \prob{\cbf}{C,L,\Tmax},\label{eq:CRCcollision_higher_chance_t}
\end{align}
By drawing a parallel between aforementioned equations and the \gls{sclf} decoding algorithm, \eqref{eq:atleastonecol} returns the probability of facing a \gls{crc} collision during its decoding in poor channel condition.
\autoref{tab:collision} returns probability of collisions for various \gls{crc} lengths, number of trials $\Tmax$ and $L=2$.
The probability increases with greater $\Tmax$ and smaller \gls{crc} lengths, as shown with \eqref{eq:probC1C2} and \eqref{eq:CRCcollision_higher_chance_t}.

In order to estimate the probability of collision in \gls{psclf}, the number of partitions, each protecting by a \gls{crc} code, should also be taken into account.
Next, the probability of having at least one \gls{crc} collision during the decoding of $P$ partitions is given.
In partitioned polar codes, the \gls{crc} structure $\mathcal{C}=\{C_1,\dots, C_p\}$ refers to the size of the $P$ \gls{crc} codes.
In \gls{psclf}, up to $\Tmax$ trials are performed, each of the trials returning $L$ candidates.
Given $\mathcal{C}$, the probability of having at least one \gls{crc} collision in one of the partitions is
\begin{align}
    \prob{\cbf}{\mathcal{C},L,\Tmax}&=1-\prod_{p=1}^{|\mathcal{C}|}\prob{\overline{\cbf}}{C_p,L,\Tmax},\label{eq:collision_w_partition}
\end{align}
where $\prob{\overline{\cbf}}{C_p,L,\Tmax}$ corresponds to the probability of having no \gls{crc} collisions with $L$ generated bit sequences for $\Tmax$ successive trials when the \gls{crc} code is of size $C_p$.
In particular, if the \gls{crc} structure is identical for all partitions, i.e., the same number of \gls{crc} bits $C$ is used in all of the $P$ partitions, \eqref{eq:collision_w_partition} can be written as
\begin{align}
    \prob{\cbf}{\mathcal{C},L,\Tmax}&=1-\left(\prob{\overline{\cbf}}{C}\right)^{L\times\Tmax\times P}.
\end{align}
As for increasing the list size $L$ \eqref{eq:CRCcollision_higher_chance_L} or increasing the number of trials $\Tmax$ \eqref{eq:CRCcollision_higher_chance_t}, having partitions increase the probability of \gls{crc} collisions, i.e.,
\begin{align}
    \prob{\cbf}{\mathcal{C},L,\Tmax}>\prob{\cbf}{C,L,\Tmax}.
\end{align}

\begin{table*}[]
    \centering
    \begin{tabular}{c|c|c|c||c|c|c}
         $C$ & $L$ & $\Tmax$ & $\prob{\cbf}{C,L,\Tmax}$ \eqref{eq:atleastonecol}& $\mathcal{C}$& $\prob{\cbf_P}{\mathcal{C},L,\Tmax}$ \eqref{eq:redpath}&$\prob{\ebf}{}$ \eqref{eq:ETeq}\\
         \hline 
         &&20&$9.3e-9$&&$9.3e-9$&0 \eqref{eq:ET_P1}\\
         32&2&50&$2.3e-8$&$\{32\}$&$2.3e-8$&0 \eqref{eq:ET_P1}\\
         &&300&$1.4e-7$&&$1.4e-7$&0 \eqref{eq:ET_P1}\\
         \hline 
         &&20&$6.1e-4$&&$3.7e-7$&0.999\\
         16&2&50&0.0015&$\{16,16\}$&$2.3e-6$&0.998\\
         &&300&0.0091&&$8.3e-5$&0.991\\
         \hline
         &&20&$\mathbf{0.145}$&&$4.4e-4$&0.997\\
         8&2&50&$\mathbf{0.32}$&$\{8,8,8,8\}$&$\mathbf{0.011}$&0.966\\
         &&300&$\mathbf{0.90}$&&$\mathbf{0.6692}$&0.26\\
         \hline
         &&20&&&$3.77e-4$&0.980\\
         &2&50&N/A&$\{7,7,7,11\}$&$\mathbf{0.0077}$&0.839\\
         &&300&&&$\mathbf{0.24}$&0.0269\\
         \hline
         &&20&&&$1.06e-4$&0.999\\
         &2&50&N/A&$\{3,11,10,8\}$&$\mathbf{0.0014}$&0.996\\
         &&300&&&$\mathbf{0.102}$&0.883\\
    \end{tabular}
    \caption{Probability of having at least one collision SCLF and PSCLF for various \gls{crc} structures. Bit sequences are considered to be randomly generated by the algorithms. Boldface probabilities are simulated and depicted in \autoref{fig:collision_proba}. }
    \label{tab:collision}
\end{table*}
\subsection{Why PSCLF Induces Many CRC Collisions ?}
As discussed in the previous section, \gls{crc} collisions happen more often with small \gls{crc} codes rather than long \gls{crc} codes \eqref{eq:crc_collision}.
Since the total number of \gls{crc} bits is fixed, smaller \gls{crc} codes are used during the encoding of partitioned polar codes.
During the decoding, it leads to an higher probability of \gls{crc} collisions in each of the partitions \eqref{eq:probC1C2}.
Moreover, our proposed decoding algorithm returns $L$ candidates during the \gls{crc} check increasing the probability of \gls{crc} collisions \eqref{eq:CRCcollision_higher_chance_L} and perform up to $\Tmax$ trials increasing furthermore the collision probability \eqref{eq:CRCcollision_higher_chance_t}.
Hence, \gls{psclf} decoding algorithm combines all properties to increase the probability of \gls{crc} collisions.

The \gls{crc} collisions are very problematic as it will pursue the decoding through the next partition despite not being a valid codeword candidate.
In the event where all \gls{crc} pass while having at least one \gls{crc} collision, the decoding is considered successful despite being invalid. 
Given that the candidates are considered to be randomly generated, each of the \gls{crc} is actually invalid.
The event of having $P$ \gls{crc} collisions is noted $\cbf_P$ and is represented as the red path in \autoref{fig:tree_diagram}. 
Its probability is estimated  to be 
\begin{align}
    \prob{\cbf_P}{\mathcal{C},L,\Tmax} &= \prod_{C\in\mathcal{C}}\prob{\cbf}{C,L,\Tmax},\label{eq:redpath}\\
    \prob{\cbf_P}{\mathcal{C},L,\Tmax} &= \prod_{C\in\mathcal{C}}\left(1-\prob{\overline{\cbf}}{C,L,\Tmax}\right).
\end{align}
An high probability for the red path will lead to validate some transmission, despite allocating a large number of \gls{crc} bits in total.

\autoref{tab:collision} returns the probability for various \gls{crc} structures with $32$ \gls{crc} bits in total, $L=2$, and $\Tmax=\{20,50,300\}$.
The probabilities shown in \autoref{tab:collision} are derived from aforementioned equations such as \eqref{eq:atleastonecol} and \eqref{eq:redpath}
For $\Tmax=300$ and $\mathcal{C}=\{8,8,8,8\}$, the probability of passing the $P$ \gls{crc} is $\prob{\cbf_P}{\mathcal{C},L,\Tmax}=0.6692$.
A more complex \gls{crc} structure reduces this probability as seen with $\mathcal{C}=\{7,7,7,11\}$ and $\mathcal{C}=\{3,11,10,8\}$. caused by the longer \gls{crc} codes.
For the first structure, being used in \cite{PPC}, the probability is now of $\prob{\cbf_P}{\mathcal{C},L,\Tmax}=0.24$ while for the second structure, used in \cite{CRC_PSCL_virtual}, the probability is  $\prob{\cbf_P}{\mathcal{C},L,\Tmax}=0.102$.
These probabilities were also verified by simulations.

\begin{figure}
    \centering
    \resizebox{0.99\columnwidth}{!}{\usetikzlibrary{spy}
\begin{tikzpicture}[spy using outlines={circle, magnification=2, connect spies}]
  \pgfplotsset{
    label style = {font=\fontsize{10pt}{8.2}\selectfont},
    tick label style = {font=\fontsize{10pt}{8.2}\selectfont}
  }

   \begin{axis}[
    width=\columnwidth,
    height=0.65\columnwidth,
    xmin=20, xmax=300,
    xlabel={$\Tmax$},
    xlabel style={yshift=0.4em},
    ymin=-0.2, ymax=1,
    ytick={0,0.25,0.5,0.75,1},
    ylabel style={yshift=-0.8em},
    ylabel={$\prob{\cbf}{}$},
    yminorticks, xmajorgrids,
    ymajorgrids, yminorgrids,
    legend style={at={(0.5,1.04)},anchor=base},
    legend style={legend columns=3, font=\scriptsize, row sep=-1.5mm},
    legend style={fill=white, fill opacity=1, draw opacity=1,text opacity=1}, 
    legend style={inner xsep=1pt, inner ysep=-2pt}, 
    mark size=1.6pt, mark options=solid,
    ]   
   \addplot[color=matlab1, mark=o,  line width=1pt,mark size=1.1pt]    table[x=T,y=PCLTmax]{proba_data.txt};
    \addplot[ color=matlab1, dashed, mark=o,line width=1pt, mark size=1.1pt]    table[x=T,y=PCLTmaxcomputed]{proba_data.txt};

    \addplot[color=matlab2, mark=x,  line width=1pt,mark size=1.1pt]    table[x=T,y=P_partition]{proba_data.txt};
    \addplot[ color=matlab2, dashed, mark=x,line width=1pt, mark size=1.1pt]    table[x=T,y=P_partitioncomputed]{proba_data.txt};

        \addplot[color=matlab3, mark=s,  line width=1pt,mark size=1.1pt]    table[x=T,y=P77711]{proba_data.txt};
    \addplot[ color=matlab3, dashed, mark=s,line width=1pt, mark size=1.1pt]    table[x=T,y=P77711computed]{proba_data.txt};

            \addplot[color=matlab4, mark=pentagon,  line width=1pt,mark size=1.1pt]    table[x=T,y=P311108]{proba_data.txt};
    \addplot[ color=matlab4, dashed, mark=pentagon,line width=1pt, mark size=1.1pt]    table[x=T,y=P311108computed]{proba_data.txt};
    \node (sc256) at (axis cs:100,0.6) [rotate=15, matlab1] {\scriptsize $\prob{\cbf}{8,2,\Tmax}$}; 
    \node (sc256) at (axis cs:125,0.35) [rotate=17, matlab2] {\scriptsize $\prob{\cbf_4}{\{8,8,8,8\},2,\Tmax}$}; 
    \node (sc256) at (axis cs:220,0.25) [rotate=7, matlab3] {\scriptsize $\prob{\cbf_4}{\{7,7,7,11\},2,\Tmax}$}; 
    \node (sc256) at (axis cs:220,-0.05) [rotate=2, matlab4] {\scriptsize $\prob{\cbf_4}{\{3,11,10,8\},2,\Tmax}$}; 
      \end{axis}

\end{tikzpicture}%
}
    \caption{Probabilities \eqref{eq:atleastonecol} and \eqref{eq:redpath} (solid lines) and simulated probabilities (dashed lines) of collisions under \gls{sclf} and \gls{psclf} decoding algorithm in poor channel condition.}
    \label{fig:collision_proba}
\end{figure}
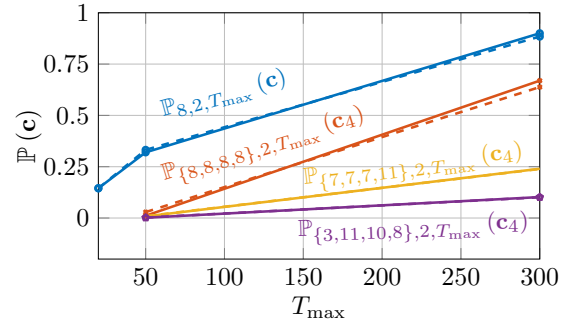

\autoref{fig:collision_proba} depicts boldfaced probabilities in \autoref{tab:collision} and their simulated values for $10000$ frames tried in poor channel condition, i.e., leading to $10000$ errors.
The simulation confirms the equations as the dashed curves, representing the simulated probabilities, and the solid curves, representing the computed probabilities are similar for different values of $\Tmax$.

\subsection{Early-termination in PSCLF}\label{subsec:et_psclf}
\begin{figure}
    \centering
    \resizebox{\columnwidth}{!}{\begin{tikzpicture}
    \coordinate (O) at (0,0);

    \coordinate (A) at (3,-2);
    \coordinate (B) at (3,2);
    \coordinate (BA) at (6,0.5);
    \coordinate (BB) at (6,3.5);
    \coordinate (BA) at (6,0.5);
    \coordinate (BB) at (6,3.5);
    \coordinate (BBA) at (9,2.5);
    \coordinate (BBB) at (9,4.5);
    \coordinate (BBBA) at (12,3.5);
    \coordinate (BBBB) at (12,5.5);

    \draw[blue, thick,dashed] (O) .. controls (1,-1) and (2,-1.5) .. (A);
    \draw[red, thick] (O) .. controls (1,1) and (2,1.5) .. (B);
    \draw[red, thick] (B) .. controls (4,3) and (5,3.5) .. (BB);
    \draw[blue, thick,dashed] (B) .. controls (4,1) and (5,0.5) .. (BA);
    \draw[red, thick] (BB) .. controls (7,4) and (8,4.5) .. (BBB);
    \draw[blue, thick,dashed] (BB) .. controls (7,3) and (8,2.5) .. (BBA);
    \draw[red, thick] (BBB) .. controls (10,5) and (11,5.5) .. (BBBB);
    \draw[blue, thick] (BBB) .. controls (10,4) and (11,3.5) .. (BBBA);
    
\node[above, red,rotate=30] at (1.5,1.2) {\Large $\prob{\cbf}{C_1,L,\Tmax}$};
\node[below, blue,rotate=-30] at (1.5,-1.2) {\Large $\prob{\overline{\cbf}}{C_1,L,\Tmax}$};
\node[below, blue,rotate=-30] at (4.5,0.8) {\Large $\prob{\overline{\cbf}}{C_2,L,\Tmax}$};
\node[below, blue,rotate=-15] at (7.5,2.8) {\Large $\prob{\overline{\cbf}}{C_3,L,\Tmax}$};
\node[below, blue,rotate=-15] at (10.5,3.8) {\Large $\prob{\overline{\cbf}}{C_4,L,\Tmax}$};
    
    \fill[draw=black,fill=black] (O) circle (2pt);
    \fill[draw=black,fill=black] (A) circle (2pt);
    \fill[draw=black,fill=black] (B) circle (2pt);
    \fill[draw=black,fill=black] (BA) circle (2pt);
    \fill[draw=black,fill=black] (BB) circle (2pt);
    \fill[draw=black,fill=black] (BBA) circle (2pt);
    \fill[draw=black,fill=black] (BBB) circle (2pt);
    \fill[draw=black,fill=black] (BBBA) circle (2pt);
    \fill[draw=black,fill=black] (BBBB) circle (2pt);
\end{tikzpicture}}
    \caption{Tree diagram of all possible PSCLF decoding paths for $P=4$. Dashed blue edges correspond to paths finishing with early-termination after failing $\Tmax$ trials. Red edges correspond to the event where the \gls{crc} has been checked during the decoding of the partition.}
    \label{fig:tree_diagram}
\end{figure}
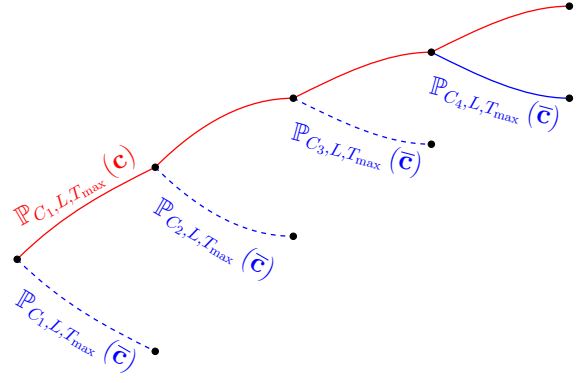

Early-termination in a decoding algorithm is the property to stop the decoding prior the end, permitting a reduction of the execution time with respect to the decoding latency.
Moreover, it also permits to reduce the average complexity of the decoding algorithm.
\gls{psclf} is able to perform early-termination thanks to the intermediate \gls{crc}.
Namely, early-termination occurs in \gls{psclf} if in the $p^{th}$ partition, with $1\leq p<P$, the \gls{crc} fails for all $L$ paths and all $\Tmax$ trials.
In \autoref{fig:tree_diagram}, the early-termination paths are represented in dashed blue.

Early-termination is a key aspect of \gls{psclf} to reduce  the decoding complexity in the low \gls{snr} region since early-termination is more likely to occur in poor channel condition.
Nevertheless, the \gls{psclf} algorithm also facilitates \gls{crc} collisions in this region as discussed in this section, especially for large value of $\Tmax$ or $L$.

The probability of early-termination in \gls{psclf} with parameters $\mathcal{C},\,L,$ and $\Tmax$, noted as the event $\ebf$, is
\begin{align}
    \prob{\ebf}{}=\sum_{p=1}^{P-1} \left(\prob{\overline{\cbf}}{C_p,L,\Tmax}\times\prod_{\rho=1}^{p-1}\prob{\cbf}{C_\rho,L,\Tmax}\right),\label{eq:ETeq}
\end{align}
by recapitulating the paths leading to an early-termination in \autoref{fig:tree_diagram}.
\autoref{tab:collision} returns probabilities for several \gls{crc} structures and several number of trials $\Tmax$.
It is observed that increasing $\Tmax$ reduces the probability of having an early-termination, which is not surprising since the probability of \gls{crc} collision increases, and thus starts the decoding of the next partition.
The \gls{crc} structure $\{7,7,7,11\}$ leads to a very small probability of early-termination with $\prob{\ebf}{}=0.0269$ when $\Tmax=300$. 
If $P=2$, \eqref{eq:ETeq} is simplified as
\begin{align}
    \prob{\ebf}{} = \prob{\overline{\cbf}}{C_1,L,\Tmax}.
\end{align}
If $P=1$, \eqref{eq:ETeq} is simplified as
\begin{align}
    \prob{\ebf}{}=0.\label{eq:ET_P1}
\end{align}
This particular case corresponds to \gls{sclf} decoding algorithm, which does not allow early-termination.

\section{Tailored Design of Partitioned Polar Codes for PSCLF Decoding Algorithm} \label{sec:designpartition}
This section discusses the partition design for codes being decoded by \gls{psclf}.
Namely, we propose to construct the partitions on the basis of the decoding capability of \gls{scl} decoder, i.e., the baseline decoder. 
This section was presented in the conference paper \cite{PPC}.
\subsection{Location of First Error in SCL}\label{sec:cdf}
Next, we denote by $X$ the random variable describing the first error in \gls{scl} for a polar code defined by $\Ical$.
The first error in \gls{scl} corresponds to the first index $i\in[N]$ where $\forall l\in[0,L-1], \mathbf{\hat{u}}_0^i[l]\neq\mathbf{u}_0^i$, i.e., all paths have diverged.
If $i\in\Fcal$, no paths duplication is performed such that the first error cannot occur. 
It cannot also occur when all decoding paths are covered, namely if $i\in\{i_1,\dots,i_{\log_2(L)}\}$.
Hence, the events of $X$ are only possible in $\{i_{\log_2(L)+1},\dots,i_{K+C}\}\triangleq\Icalsort$.
The \gls{cdf} of $X$ is denoted $F(k)$, and is defined as 
\begin{align}
    F(k) = \mathbb{P}(X \leq k)=\sum_{i=0}^k \mathbb{P}(X=i)\,,
\end{align}
with $\mathbb{P}(X=i)=0$ if $i\in\Fcal$ or $i\in\{i_1,\dots,i_{\log_2(L)}\}$.

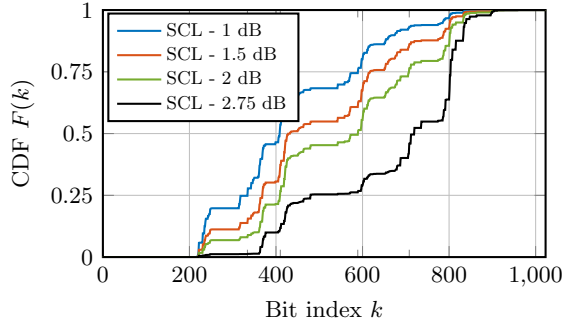
\begin{figure}
    \centering
    \vspace{2pt}
    \resizebox{.99\columnwidth}{!}{\begin{tikzpicture}
  \pgfplotsset{
    label style = {font=\fontsize{9pt}{7.2}\selectfont},
    tick label style = {font=\fontsize{9pt}{7.2}\selectfont}
  }
   
   \begin{axis}[%
    width=\columnwidth,
    height=0.65\columnwidth,
    xmin=0, xmax=1023,
    minor xtick={334,410,589,708},
    xlabel={Bit index $k$},
    xlabel style={yshift=0.4em},
    ymin=0, ymax=1,
    ytick={0,0.25,0.5,0.75,1},
    ylabel style={yshift=-0.4em},
    ylabel={CDF $F(k)$},
    xlabel style={yshift=-0.2em},
    yminorticks, xmajorgrids,
    ymajorgrids, yminorgrids,
    legend style={at={(0.01,0.99)},anchor=north west},
    legend style={nodes={scale=0.8}, font=\small},
     legend style={legend columns=1},
     legend cell align={left},
    mark size=1.6pt, line width=0.8pt, mark options=solid,
    ] 


 \addplot [solid, color=matlab1, mark phase=0]
 table[x=xdata,y=ydata]{CDF_L4_1024_512_32_1dB.txt};
 \addlegendentry{\gls{scl} - 1 dB}

  \addplot [solid, color=matlab2, mark phase=0]
 table[x=xdata,y=ydata]{CDF_L4_1024_512_32_1.5dB.txt};
 \addlegendentry{\gls{scl} - 1.5 dB}

   \addplot [solid, color=matlab5, mark phase=0]
 table[x=xdata,y=ydata]{CDF_L4_1024_512_32.txt};
 \addlegendentry{\gls{scl} - 2 dB}
    \addplot [solid, color=black, mark phase=0]
 table[x=xdata,y=ydata]{CDF_L4_1024_512_32_2.75dB_new.txt};
 \addlegendentry{\gls{scl} - 2.75 dB}
\end{axis}    
\end{tikzpicture}%
}
    \caption{\gls{cdf} $F(k)$ of $(1024,512+32)$ polar code for $L=4$ and $\text{SNR}=\{1,\,1.5,\, 2,\,2.75\}$ dB.}
    \label{fig:CDF_SNR}
\end{figure}
\autoref{fig:CDF_SNR} depicts $F(k)$ for $L=4$ of a $(1024,512+32)$ polar code at $\text{SNR}=\{1,\, 1.5,\, 2,\,2.75\}$\,dB.
As seen in \autoref{fig:CDF_SNR}, the probability that the first error occurs earlier in $\hat{u}$ increases with the noise.
For instance, the location delimiting half of the first errors, i.e., $F(k)=0.5$ is reached at $k=\{409,433,590,720\}$ for $\text{SNR}=\{1,\, 1.5,\, 2,\,2.75\}$\,dB, respectively.

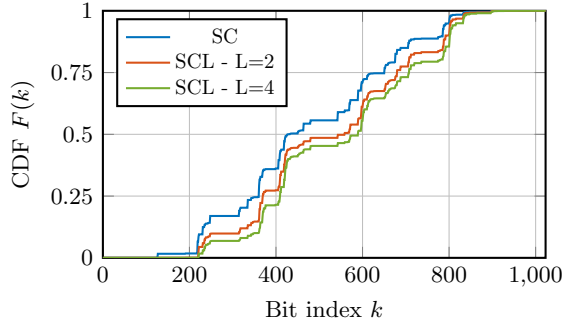
\begin{figure}[t]
    \centering
    \vspace{2pt}
    \resizebox{.99\columnwidth}{!}{\begin{tikzpicture}
  \pgfplotsset{
    label style = {font=\fontsize{9pt}{7.2}\selectfont},
    tick label style = {font=\fontsize{9pt}{7.2}\selectfont}
  }
   
   \begin{axis}[%
    width=\columnwidth,
    height=0.65\columnwidth,
    xmin=0, xmax=1023,
    xlabel={Bit index $k$},
    xlabel style={yshift=0.4em},
    ymin=0, ymax=1,
    ytick={0,0.25,0.5,0.75,1},
    ylabel style={yshift=-0.4em},
    ylabel={CDF $F(k)$},
    xlabel style={yshift=-0.2em},
    yminorticks, xmajorgrids,
    ymajorgrids, yminorgrids,
    legend pos=north west,
    legend style={nodes={scale=0.8}},
     legend style={legend columns=1},
    mark size=1.6pt, line width=0.8pt, mark options=solid,
    ] 


 \addplot [solid, color=matlab1, mark phase=0]
 table[x=xdata,y=ydata]{CDF_1_1024_512_32.txt};
 \addlegendentry{\gls{sc}}

  \addplot [solid, color=matlab2, mark phase=0]
 table[x=xdata,y=ydata]{CDF_L2_1024_512_32.txt};
 \addlegendentry{\gls{scl} - L=2}

   \addplot [solid, color=matlab5, mark phase=0]
 table[x=xdata,y=ydata]{CDF_L4_1024_512_32.txt};
 \addlegendentry{\gls{scl} - L=4}
\end{axis}    
\end{tikzpicture}%
}
    \caption{\gls{cdf} $F(k)$ of $(1024,512+32)$ polar code for various list sizes $L$ at $\text{SNR}=2$ dB.}
    \label{fig:CDF_L}
\end{figure}
\autoref{fig:CDF_L} depicts $F(k)$ of a $(1024,512+32)$ polar code for various list sizes $L=\{1,2,4\}$ and $\text{SNR}=2$\,dB. 
As seen in \autoref{fig:CDF_L}, as the list size $L$ grows, the probability to have a first error early in the frame decreases.
It is explained by the shift to the end of the first index of $\Icalsort$ and the improved error capability of \gls{scl} with a larger list size.
The location delimiting a quarter of the first errors, i.e., $F(k)= 0.25$ is reached at $k=\{361, 370 , 410\}$ for $L=\{1,2,4\}$, respectively.
\subsection{Proposed Partition Design}\label{sec:cdf_design}
The proposed design of $\mu$ is to uniformly distribute the partitions according to the \gls{cdf} $F(k)$.
Namely, $\forall p\in\{1,\dots,P\}, \mu_p$ verifies
\begin{align}
    F(\mu_p-1)<\frac{p}{P}\leq F(\mu_p).\label{eq:constructioncdf}
\end{align}
A similar design for \gls{pscf} was carried out by using the decoding behavior of \gls{sc} \cite{PSCF}.
Since the proposed \gls{psclf} algorithm uses \gls{scl} as its core decoder, the \glspl{cdf} are based on the decoding behavior of \gls{scl}.
Given \autoref{fig:CDF_SNR} and for a simulation using \gls{scl} with $L=4$, the three sets $\mu$ are
\begin{align}
    \mu&=\{335,409,589,1023\},\label{eq:mu_1dB}\\
    \mu&=\{410,590,708,1023\},\label{eq:mu_2dB}\\
    \mu&=\{423,720,804,1023\},\label{eq:mu_275dB}       
\end{align}
for a design SNR at $1$, $2$, and $2.75$ dB.

The number of non-frozen bits in each partition is based on $\mu$ and the information set $\Ical$ of the polar code.
For these particular examples, the number of non-frozen bits $\mathcal{K}=\{s_1,s_2,s_3,s_4\}$ inside each partition is $\mathcal{K}=\{28,32,100,384\}$ if $\mu$ \eqref{eq:mu_1dB}, $\mathcal{K}=\{61,100,84,299\}$ if $\mu$ \eqref{eq:mu_2dB}, and $\mathcal{K}=\{71,107,220,204\}$ if $\mu$ \eqref{eq:mu_275dB}.
All partitions are protected by a \gls{crc} code and given  the sets $\mathcal{K}$, the number of non-frozen bits fluctuates heavily with the partitions.
\subsection{Summary of All Partitions}
The partitions designed for all codes are given below. 
Partitions are designed according to the aforementioned method, where the \gls{cdf} was analyzed for a practical frame-error-rate region of $10^{-3}$ as in \cite{PSCF}.
Moreover, two additional methods $\mu_{divN}$ and $\mu_{divK}$ are also shown.
The first design creates partitions of same size, while the second design allocates the same number of information bits per partition.
\begin{table*}[t]
    \centering
    \begin{tabular}{c|c|c|c|c|c}
         N& K + C& Design SNR [dB]& $\mu$ - $\mathcal{K}$ & $\mu_{\text{divN}}$ - $\mathcal{K}_{\text{divN}}$ \eqref{eq:defdivN}& $\mu_{\text{divK}}$ - $\mathcal{K}_\text{divK}$ \eqref{eq:defdivK}  \\
         \hline
         \multicolumn{6}{c}{Code designed with Density Evolution and $\mu$ found with \eqref{eq:constructioncdf}}\\
         \hline
         \multirow{6}{*}{1024}& \multirow{2}{*}{256+32} & \multirow{2}{*}{-1.5}  & 490 730 861 & N/A & 730 867 950\\
         &  &  &  24 48 67& N/A & 72 144 216\\
         \cline{2-6}
         & \multirow{2}{*}{512+32} & \multirow{2}{*}{2.75}  &423 720 804 &255 511 767  & 496 735 887 \\ 
         & & &71 107 220 & 24 128 154  & 136 272 408 \\ 
         \cline{2-6}
         & \multirow{2}{*}{768+32} & \multirow{2}{*}{5.25}  & 210 402 586 &255 511 767  & 394 620 823\\ 
         & &   & 67 141 158 &102 215 228  & 200 400 600\\ 
         \hline
    \end{tabular}
    \caption{Code parameters, design SNR for $\Ical$ and partitions used in this paper, the last index of $\mu_P=N-1$ \eqref{eq:muPval} ($\mathcal{K}_P=K+C$ \eqref{eq:defdivK}) is omitted.}
    \label{tab:partition_used}
\end{table*}
\subsection{CRC Structure of the Partitioned Polar Codes}
With the exception of \cite{CRC_PSCL_virtual}, partitioned decoding of polar codes is usually carried out with a fixed number of \gls{crc} bits per partitions. 
Defining the set of \gls{crc}-code size per partition as $\mathcal{C}=\{C_1,\dots,C_P\}$, most works have $C_1=\dots=C_P$.
In \cite{CRC_PSCL_virtual}, the \gls{crc} structure was designed according to the capacity of the sub-channels given a set $\mu$. 
As a consequence, this method designs the \gls{crc} structure according to the length of the partitions as well as the reliability of the polar code.
However, the design of $\mu$ was not discussed and $\mu=\{255,511,767,1023\}$ was chosen with $\mathcal{K}=\{20,123,156,245\}$ non-frozen bits in each partition.
The resulting \gls{crc} structure is $\mathcal{C}_1=\{3,11,10,8\}$ \cite{CRC_PSCL_virtual}.

\begin{table}[t!]
    \vspace{10pt}
    \centering
    \begin{tabular}{|c|>{\centering}p{1.2cm}|>{\centering}p{1.2cm}|>{\centering}p{1.2cm}|c|}
        \cline{2-5}
         \multicolumn{1}{c|}{} &\multicolumn{4}{c|}{Partition $p$}  \\
        \cline{2-5}
         \multicolumn{1}{c|}{} &1&2&3&4  \\
         \hline 
         1&0.53&0.23&0.17&0.07\\
         1.5&0.37&0.25&0.24&0.14\\
         2&0.25&0.25&0.25&0.25\\
         2.75&0.09&0.14&0.21&0.56\\
         \hline
         \multirow{2}{*}{$\nicefrac{E_b}{N_0}$ dB}&\multicolumn{4}{c|}{Probability $\mathbb{P}(e_p)$ that the }\\
         &\multicolumn{4}{c|}{first error occurs in partition $p$.}\\
         \hline
    \end{tabular}
    \caption{Probabilities that the first error occurs in each partition, the set $\mu$ is designed at $2$ dB \eqref{eq:mu_2dB}.} 
    \label{tab:prob_error}
\end{table}
The proposed \gls{crc} structure $\mathcal{C}$ in \cite{PPC} is designed according to the probability of error in each partition and the decoding behavior of \gls{scl}.
In a partition, an error is either caused by having no paths passing the \gls{crc} or having one path passing the \gls{crc} but being a false positive, i.e, a \gls{crc} collision.
The former may end up getting corrected by a decoding-path flip.
The latter, however, cannot be corrected by \gls{psclf} since only wrong paths will end up being forwarded to the next partition.
Hence, \gls{crc} collisions should be reduced and the proposed \gls{crc} structure does so by taking the probability of errors in each partition into account.

Given \autoref{fig:CDF_SNR}, at low $\frac{E_b}{N_0}$ values, a decoding error is mostly due to an error happening in the earlier indices of a codeword  while at high $\frac{E_b}{N_0}$ values a decoding error is mostly occurring in the latter indices.
\autoref{tab:prob_error} shows the probability that the first error occurs for each partition, the set $\mu$ is designed at $2$ dB \eqref{eq:mu_2dB}.
For a given partition $p$ and $\frac{E_b}{N_0}$, the probability $\mathbb{P}(e_p)$ that the first error occurs can be retrieved from \autoref{fig:CDF_SNR}, i.e.,
\begin{align}
    \mathbb{P}(e_p)=F(\mu_p)-F(\mu_{p-1}+1).
\end{align}
The number of \gls{crc} bits are allocated according to the target \gls{snr}. 
If the design is performed at $\frac{E_b}{N_0}=1$ dB, more \gls{crc} bits should be allocated in the first partition since $P(e_1)=53$\%.
If the design is performed $\frac{E_b}{N_0}=2.75$ dB, more \gls{crc} bits should be allocated towards the last partition since $p(e_4)=56$\%.

\section{Simulation Results}\label{sec:sim_results}
All simulations are performed over the \gls{awgn} channel using the \gls{bpsk} modulation.
The information on the design SNR to compute the reliability order is given in Table \ref{tab:partition_used}, as well as the partitions used.
With respect to \cite{PPC}, this extended paper provides simulation results for $L=2$ instead of $L=4$.
The reason is to limit the impact of the \gls{crc} collisions, increased by the larger number of trials $\Tmax$ required with  $\omega>1$.
Moreover, several code rates and decoding order $\omega$ are investigated, while \cite{PPC} provided results for rate $\frac{1}{2}$ and $\omega=1$ only.
We set $T_{\max}=\{20,50,300\}$ for $\omega=\{1,2,3\}$.
\subsection{Impact of the Decoding Order $\omega$}
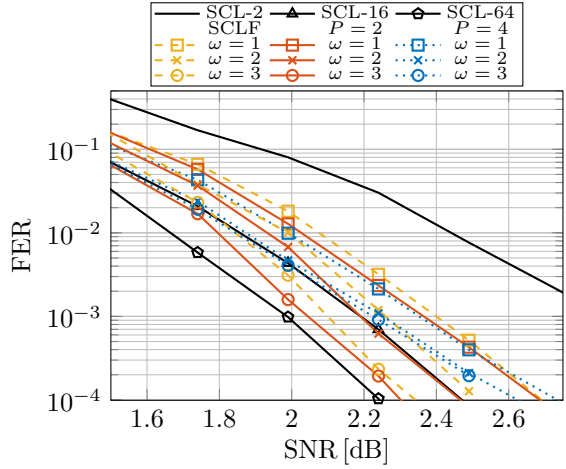
\begin{figure}[t!]
    \centering
    \vspace{4pt}
    \resizebox{.99\columnwidth}{!}{\usetikzlibrary{spy}
\begin{tikzpicture}[spy using outlines={circle, magnification=2, connect spies}]
  \pgfplotsset{
    label style = {font=\fontsize{10pt}{8.2}\selectfont},
    tick label style = {font=\fontsize{10pt}{8.2}\selectfont}
  }

   \begin{semilogyaxis}[
    width=\columnwidth,
    height=0.75\columnwidth,
    xmin=1.5, xmax=2.75,
    xlabel={$\text{SNR}\,[\mathrm{dB}]$},
    xlabel style={yshift=0.4em},
    ymin=9.9e-5, ymax=0.5,
    ylabel style={yshift=-0.1em},
    ylabel={FER},
    yminorticks, xmajorgrids,
    ymajorgrids, yminorgrids,
    legend style={at={(0.5,1.05)},anchor=base},
    legend style={legend columns=3, font=\footnotesize, row sep=-1.5mm},
    legend style={fill=white, fill opacity=1, draw opacity=1,text opacity=1}, 
    legend style={inner xsep=1pt, inner ysep=-1pt}, 
    mark size=1.6pt, mark options=solid,
    ]   
       \addplot[color=black,line width=0.8pt, mark size=2.1pt]
    table[x=SNR,y=SCL2]{SCL_1024_512.tex};
    \addlegendentry{SCL-2}
    \addplot[color=black, mark=triangle,line width=0.8pt, mark size=2.1pt]
    table[x=SNR,y=SCL16]{SCL_1024_512.tex};
    \addlegendentry{SCL-16}
        \addplot[color=black, mark=pentagon,line width=0.8pt, mark size=2.1pt]
    table[x=SNR,y=SCL64]{SCL_1024_512.tex};
    \addlegendentry{SCL-64}

    \addlegendimage{empty legend}
    \addlegendentry{SCLF}
    \addlegendimage{empty legend}
    \addlegendentry{$P=2$}
    \addlegendimage{empty legend}
    \addlegendentry{$P=4$}
   \addplot[color=matlab3, dashed,mark=square, line width=0.8pt, mark size=2.1pt]
    table[x=snr,y=FER]{1024_512_20.txt};
    \addlegendentry{$\omega=1$}      
   
   \addplot[color=matlab2,mark=square, line width=0.8pt, mark size=2.1pt]
    table[x=snr,y=FER]{1024_512_20_2.txt};
    \addlegendentry{$\omega=1$}
       \addplot[color=matlab1,dotted, mark=square, line width=0.8pt, mark size=2.1pt]
    table[x=snr,y=FER]{1024_512_20_4.txt};
    \addlegendentry{$\omega=1$}
   \addplot[color=matlab3,dashed, mark=x, line width=0.8pt, mark size=2.1pt]
    table[x=snr,y=FER]{1024_512_50.txt};
    \addlegendentry{$\omega=2$}
   \addplot[color=matlab2, mark=x, line width=0.8pt, mark size=2.1pt]
    table[x=snr,y=FER]{1024_512_50_2.txt};
    \addlegendentry{$\omega=2$}
    
   \addplot[color=matlab1,dotted, mark=x, line width=0.8pt, mark size=2.1pt]
    table[x=snr,y=FER]{1024_512_50_4.txt};
    \addlegendentry{$\omega=2$}
    \addplot[color=matlab3,dashed, mark=o, line width=0.8pt, mark size=2.1pt]    table[x=snr,y=FER]{1024_512_300.txt};    \addlegendentry{$\omega=3$}
    \addplot[color=matlab2,mark=o, line width=0.8pt, mark size=2.1pt]    table[x=snr,y=FER]{1024_512_300_2.txt};    \addlegendentry{$\omega=3$}
    
    \addplot[color=matlab1, dotted, mark=o, line width=0.8pt, mark size=2.1pt]    table[x=snr,y=FER]{1024_512_300_4.txt};    \addlegendentry{$\omega=3$}
    
  \end{semilogyaxis}

\end{tikzpicture}%
}
    \caption{\Gls{fer} with decoding order $\omega=\{1,2,3\}$, number of partitions $P=\{1,2,4\}$ for $N=1024$, $K=512$, and $C=32$.}
    \label{fig:FER_omega}
\end{figure}

In this subsection, the impact of the decoding order $\omega$ is shown with $P=\{1,2,4\}$ partitions.
For this example, code parameter with $(N,K)=(1024,512)$ is used.
$32$ \gls{crc} bits are used, with a \gls{crc} structure evenly distributing the \gls{crc} bits among the partitions, i.e., $\mathcal{C}=\{8,8,8,8\}$.
Moreover, the partition  is designed as $\mu$ described in \autoref{tab:partition_used}.
\autoref{fig:FER_omega} depicts the error-correction performance.

For $\omega=1$, depicted with squared markers, \gls{psclf} provides greater performance with respect to \gls{sclf} ($P=1$) for $\text{FER}>2\cdot10^{-4}$.
The gain reaches $0.08$\,dB for $P=4$ at $\text{FER}=10^{-2}$, leading to a gain of $0.43$\,dB with respect to SCL-2, the baseline algorithm of our proposed algorithm.

For $\omega=2$, depicted with cross markers, using $P=4$ partitions improves performance for $\text{FER}>3\cdot10^{-3}$, reaching a $0.13$\,dB gain with respect to \gls{sclf} at $\text{FER}=10^{-2}$.
Using $P=2$ permits a coding gain for all depicted \glspl{fer}.
The performance of the proposed approach is close or identical to that of SCL-16.

For $\omega=3$, depicted with circle markers, the performance marginally improves compared to $\omega=2$ with $P=4$. 
As discussed in \autoref{sec:collisions}, this is caused by the \gls{crc} collisions. With $\omega=3$, the algorithm requires a greater number of additional trials $\Tmax$.
Having $P=2$ permits to obtain a gain with respect \gls{sclf} and SCL-16. 
With $P=2$, a $0.05$\,dB gain is observed at $\text{FER}=10^{-2}$.

In conclusion, for all values of $P$, increasing the decoding order $\omega$ and maximum number of trials $\Tmax$ improves the decoding performance, as expected.
Nevertheless, having  $P=4$ leads to worse performance with respect to $P=2$ at small \gls{fer}.
As mentioned in \cite{PPC}, this is caused by \gls{crc} collisions, as fewer \gls{crc} bits can be allocated to each partition.
Moreover, the impact is greater for $\omega\geq2$, which is explained by the number of \gls{crc} collisions being greater, as the number of flipping trials $\Tmax$ also increases.
\subsection{Impact of the Partition Designs}
In \cite{PPC}, the proposed design described in \autoref{sec:designpartition} is shown to improve the error-correction performance with respect to $\mu_{divN}$.
The comparison was performed only for the code parameters $(N,K)=(1024,512)$.
The gain is shown to be greater for $P=4$ with respect to $P=2$.
Moreover, for $P=4$, the design is shown to delay the error-correction performance degradation caused by the \gls{crc} collisions as the \gls{fer} decreases.
Next, the comparison is performed for a different number of flips per trial, i.e., $\omega=2$ and $\Tmax=50$.
Three rates are investigated and comparisons with $\mu_{divK}$ are also provided.
The partitions used are described in \autoref{tab:partition_used}.
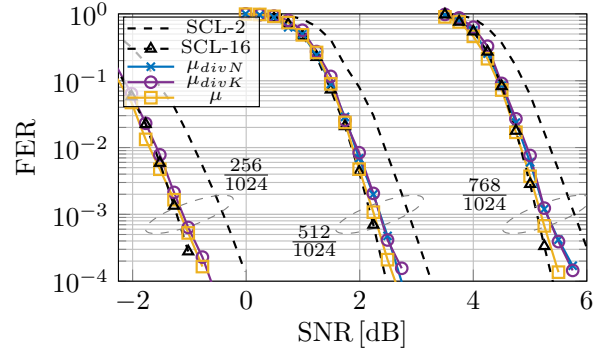
\begin{figure}
    \centering
    \resizebox{0.5\textwidth}{!}{\usetikzlibrary{spy}
\begin{tikzpicture}[spy using outlines={circle, magnification=2, connect spies}]
  \pgfplotsset{
    label style = {font=\fontsize{10pt}{8.2}\selectfont},
    tick label style = {font=\fontsize{10pt}{8.2}\selectfont}
  }

   \begin{semilogyaxis}[
    width=\columnwidth,
    height=0.66\columnwidth,
    xmin=-2.25, xmax=6,
    xlabel={$\text{SNR}\,[\mathrm{dB}]$},
    xlabel style={yshift=0.4em},
    ymin=1e-4, ymax=1,
    ytick={0.0001,0.001,0.01,0.1,1},
    ylabel style={yshift=-0.1em},
    ylabel={FER},
    yminorticks, xmajorgrids,
    ymajorgrids, yminorgrids,
    legend style={at={(0.01,0.99)},anchor=north west},
    legend style={legend columns=1, font=\footnotesize},
    legend style={fill=white, fill opacity=0.7, draw opacity=1,text opacity=1}, 
    legend style={inner xsep=1pt, inner ysep=-1pt, row sep=-1mm}, 
    mark size=1.6pt, mark options=solid,
    ]   
    \addplot[color=black, dashed,line width=0.8pt, mark size=2.1pt]
    table[x=SNR,y=SCL2]{SCL_1024_512.tex};
    \addlegendentry{SCL-2}
    \addplot[color=black, dashed, mark=triangle,line width=0.8pt, mark size=2.1pt]
    table[x=SNR,y=SCL16]{SCL_1024_512.tex};
    \addlegendentry{SCL-16}
   \addplot[color=matlab1, mark=x, line width=0.8pt, mark size=2.1pt]
    table[x=SNR,y=FERmueq]{1024_512_mu_P4_w2.tex};
    \addlegendentry{$\mu_{divN}$}      
   \addplot[color=matlab4, mark=o, line width=0.8pt, mark size=2.1pt]
    table[x=SNR,y=FERdivK]{1024_512_mu_P4_w2.tex};
    \addlegendentry{$\mu_{divK}$}      
   \addplot[color=matlab3, mark=square, line width=0.8pt, mark size=2.1pt]
    table[x=SNR,y=FER]{1024_512_mu_P4_w2.tex};
     \addlegendentry{$\mu$}  
     \addplot[color=matlab1, mark=x, line width=0.8pt, mark size=2.1pt]
    table[x=SNR,y=FERmueq]{1024_768_mu_P4_w2.tex};
   
   \addplot[color=matlab4, mark=o, line width=0.8pt, mark size=2.1pt]
    table[x=SNR,y=FERdivK]{1024_768_mu_P4_w2.tex};
   \addplot[color=matlab3, mark=square, line width=0.8pt, mark size=2.1pt]
    table[x=SNR,y=FER]{1024_768_mu_P4_w2.tex};

       \addplot[color=matlab4, mark=o, line width=0.8pt, mark size=2.1pt]
    table[x=SNR,y=FERdivK]{1024_256_mu_P4_w2.tex};
   \addplot[color=matlab3, mark=square, line width=0.8pt, mark size=2.1pt]
    table[x=SNR,y=FER]{1024_256_mu_P4_w2.tex};
       \addplot[color=black, dashed,line width=0.8pt, mark size=2.1pt]
    table[x=SNR,y=SCL2]{SCL_1024_768.tex};
    \addplot[color=black, dashed, mark=triangle,line width=0.8pt, mark size=2.1pt]
    table[x=SNR,y=SCL16]{SCL_1024_768.tex};
           \addplot[color=black, dashed,line width=0.8pt, mark size=2.1pt]
    table[x=SNR,y=SCL2]{SCL_1024_256.tex};
    \addplot[color=black, dashed, mark=triangle,line width=0.8pt, mark size=2.1pt]
    table[x=SNR,y=SCL16]{SCL_1024_256.tex};
    \draw[dashed,gray]  (axis cs:2.35,0.001) ellipse [    x radius = 20, y radius = 1.6,rotate=110];
    \node [rotate=0]at (axis cs:1.2,0.0004) {\scriptsize $\frac{512}{1024}$};
    \draw[dashed,gray]  (axis cs:-1,0.001) ellipse [    x radius = 20, y radius = 1.6,rotate=110];
    \node [rotate=0]at (axis cs:0,0.004) {\scriptsize $\frac{256}{1024}$};
    
    \draw[dashed,gray]  (axis cs:5.29,0.001) ellipse [    x radius = 20, y radius = 1.6,rotate=110];
    \node [rotate=0]at (axis cs:4.2,0.002) {\scriptsize $\frac{768}{1024}$};
  \end{semilogyaxis}

\end{tikzpicture}%
}
    \caption{Error-correction performance with various constructions of partitions $\mu$ described in \autoref{tab:partition_used}. Parameters are  $P=4$, $N=1024$, $K=\{256,512,768\}$, $\mathcal{C}=\{8,8,8,8\}$, and $\omega=2$. Error-correction performance of \gls{crc}-aided SCL for $L=\{2,16\}$ are shown as reference.}
    \label{fig:impactmudesign}
\end{figure}

\begin{figure}
    \centering
    \resizebox{0.99\columnwidth}{!}{\usetikzlibrary{spy}
\begin{tikzpicture}[spy using outlines={circle, magnification=2, connect spies}]
  \pgfplotsset{
    label style = {font=\fontsize{10pt}{8.2}\selectfont},
    tick label style = {font=\fontsize{8pt}{8.2}\selectfont}
  }

   \begin{semilogyaxis}[%
    width=\columnwidth,
    height=0.75\columnwidth,
    xmin=1e-04, xmax=1e-00,
    xlabel={Frame-error rate},
    xlabel style={yshift=0.4em},
    ymin=2000, ymax=1000000,
    x dir=reverse,
    xmode=log,
    ytick = {2624,10496, 41986,167944,671766},
    yticklabels={$\latSCL$,4$\latSCL$,16$\latSCL$,64$\latSCL$,256$\latSCL$},
    ylabel style={yshift=1.1em},
    ylabel={Avg. Exec. Time},
    xlabel style={yshift=0.2em},
    yminorticks, xmajorgrids,
    ymajorgrids, yminorgrids,
    legend style={at={(0.99,1.0)},anchor=north east},
    legend style={legend columns=2, font=\scriptsize, column sep=0mm, row sep=-1mm}, 
    legend cell align={left},
    mark size=1.8pt, mark options=solid,
    ]

  \addplot [dashed, color=black, line width=0.8pt, mark size=2.1pt]table[row sep=crcr]{
  3.2e-04   2624\\      
  1e-00   2624\\
  };  
  \addlegendentry{SCL-2}

           \addplot[color=matlab2,mark=o, line width=0.8pt, mark size=2.1pt]
   table[x=FER,y=cc]{PSCLF3_L2_1024_544_16_16_CR.txt};   
    \addlegendentry{P=2}
\addplot [color=matlab1,mark=x, line width=0.8pt, mark size=2.1pt]
  table[x=FER8888,y=avg8888]{1024_512_diff_crc_avg_exec_time.tex};
\addlegendentry{P=4}

  \addplot [color=black, line width=0.8pt, mark size=2.1pt]
  table[x=FER,y=ccnogrm]{SCLF3_LLRM_GRM_1024_544.txt};
  \addlegendentry{SCLF}   
  \addplot [color=black,dotted, line width=0.8pt, mark size=2.1pt]
  table[x=FER,y=ccllrm]{SCLF3_LLRM_GRM_1024_544.txt};
  \addlegendentry{SCLF - LLRM}

  \end{semilogyaxis}


\end{tikzpicture}}
    \caption{Average execution time for $(N,K)=(1024,512)$ with $\omega=3$.}
    \label{fig:exec_psclf}
\end{figure}
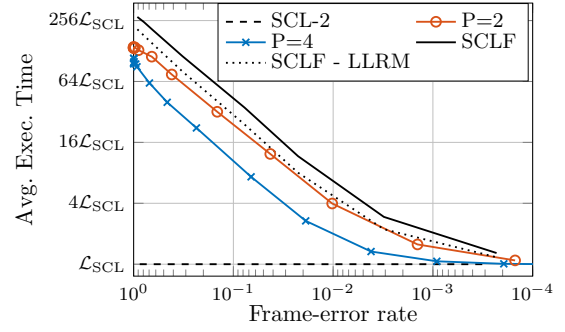

\autoref{fig:impactmudesign} depicts the error-correction performance for $P=4$.
For reference, \gls{scl} performance is shown for $L=\{2,16\}$.
For $\frac{K}{N}=\frac{256}{1024}$, the construction leading to $\mu_{divN}$ is impossible as no information bits are declared in the first partition.
For all rates, the proposed design of partitions improves the performance by approximately {$0.15$\,dB at $\text{FER}=10^{-2}$.

\subsection{Impact of the CRC Structure}\label{subsec:CRC_sim}
In \cite{PPC}, the impact of the \gls{crc} structure for PSCLF with $\omega=1$ and $\Tmax=30$ for $(1024,512+32)$ is discussed.
Namely, the following \gls{crc} structure $\mathcal{C}_2=\{7,7,7,11\}$ was proposed to improve performance at low \gls{fer} with respect to the CRC structure $\mathcal{C}=\{8,8,8,8\}$.
Indeed, it was observed at low \gls{fer} that PSCLF was facing more \gls{crc} collisions for the last partition.
Hence, reducing to 7 \gls{crc} bits did not significantly increase \gls{crc} collisions for the first three partitions, while allocating 11 CRC bits to the last partition reduced the number of CRC collisions.
This was also explained by the number of trials, being limited to $\Tmax=30$.
In this paper, the average chance of CRC collisions increases with the number of trials $\Tmax$.

\autoref{fig:FER_CRC} shows the error-correction performance for $P=4$, $(1024,512+32)$ for three \gls{crc} structures.
These parameters were chosen due to poor gain between $\omega=3$ and $\omega=2$  in \autoref{fig:FER_omega}.
While the \gls{crc} structures had an impact in \cite{PPC}, the high number of trials $\Tmax$ limits the impact of the \gls{crc} structure.
At very high \gls{fer}, the structure $\mathcal{C}_1=\{11,7,7,7\}$ has the best performance. However, it performs worse at low \gls{fer} for the same reason explained in \cite{PPC}: the errors are now concentrated in the last part of the codeword.
Hence, flipping is then performed, which may lead to \gls{crc} collisions, and thus, an error.
For the structure $\mathcal{C}_1=\{7,7,7,11\}$, the performance at low \gls{fer} is worst.
The errors are concentrated in the beginning, inducing flipping.
This increases the chance of \gls{crc} collisions which leads to pursue the decoding with a wrong candidate.
Finally, the simplest structure, dividing the \gls{crc} bits equally, looks to be the best option when the decoding order increases, as proposed in this paper.

In order to fully exploit the proposed algorithm with $\omega=3$, more \gls{crc} bits will be required to avoid the \gls{crc} collisions. However, $32$ is already a large number of \gls{crc} bits.

\begin{figure}[t!]
    \centering
    \vspace{2pt}
    \resizebox{.99\columnwidth}{!}{\begin{tikzpicture}
  \pgfplotsset{
    label style = {font=\fontsize{10pt}{8.2}\selectfont},
    tick label style = {font=\fontsize{10pt}{8.2}\selectfont}
  }

   \begin{semilogyaxis}[
    width=\columnwidth,
    height=0.75\columnwidth,
    xmin=1.0, xmax=2.75,
    xlabel={$\text{SNR}\,[\mathrm{dB}]$},
    xlabel style={yshift=0.4em},
    ymin=1e-5, ymax=0.5,
    ylabel style={yshift=-0.1em},
    ylabel={FER},
    yminorticks, xmajorgrids,
    ymajorgrids, yminorgrids,
    legend style={at={(0.01,0.01)},anchor=south west},
    legend style={legend columns=1, font=\scriptsize, row sep=-1mm},
    legend style={fill=white, fill opacity=1, draw opacity=1,text opacity=1}, 
    legend cell align={left},
    legend style={inner xsep=0.2pt, inner ysep=-1pt}, 
    mark size=1.6pt, mark options=solid,
    ]   

       \addplot[color=black, mark=diamond, line width=0.8pt, mark size=2.1pt]
table[x=xdata,y=ydata]{sclf.txt}; 
    \addlegendentry{SCLF}  

\addplot[color=matlab1,mark=x, line width=0.8pt, mark size=2.1pt]
   table[x=SNR,y=FER]{PSCLF3_L2_1024_544_8_8_8_8.txt}; 
    \addlegendentry{$\mathcal{C}=\{8,8,8,8\}$}

   \addplot[color=matlab3,mark=square, line width=0.8pt, mark size=2.1pt]
   table[x=SNR,y=FER]{PSCLF3_L2_1024_544_11_7_7_7.txt};  
    \addlegendentry{$\mathcal{C}_1=\{11,7,7,7\}$}
    
       \addplot[color=matlab4,mark=otimes, line width=0.8pt, mark size=2.1pt]
   table[x=SNR,y=FER]{PSCLF3_L2_1024_544_7_7_7_11.txt};  
    \addlegendentry{$\mathcal{C}_2=\{7,7,7,11\}$}
  \end{semilogyaxis}

\end{tikzpicture}}
    \caption{\gls{fer} of $(1024,512+32)$ with multiple \gls{crc} structures with $P=4$. }
    \label{fig:FER_CRC}
\end{figure}
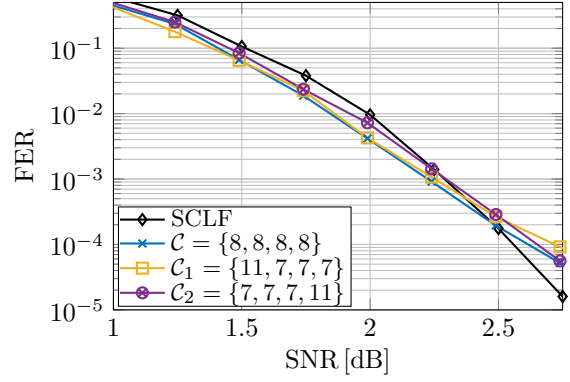

\subsection{Average Execution Time of \gls{psclf}}\label{subsec:avg_exec_time_sim}
The baseline SCL decoder has a list size of $L=2$ in PSCLF.
The number of processing elements is set to $\varphi=64$.
As described in \autoref{subsec:exec_psclf}, the average execution time $\avgPSCLF$ of PSCLF \eqref{eq:avgPSCLF} depends on the design of partitions, the average number of additional trials performed per partition, the number of processing elements $\varphi$ and the size of the information set $|\Ical|$ \eqref{eq:latSCL_all}\eqref{eq:latSCL}.

The average execution time of the proposed \gls{psclf} is also studied in \cite{PPC}.
It was shown that the average execution time of \gls{psclf} is reduced in comparison to that of \gls{sclf}, i.e., partitioned polar codes help reduce the overall decoding complexity.
The reduction increased with $P$.

Next, the average execution time is depicted for all $\omega=\{1,2,3\}$ and $P=\{2,4\}$ in Figures\,\ref{fig:exec_psclf20}, \ref{fig:exec_psclf50}, and \ref{fig:exec_psclf}.
All figures use the \gls{fer} as x-axis for a direct comparison with the error-correction performance.
A lower frame-error rate induces a better channel condition.
The latency of the SCL baseline decoder $\latSCL$ \eqref{eq:latSCL_all} is shown, in the figures, for reference.
Moreover, since the average execution time of PSCLF heavily depends on the latency of SCL \eqref{eq:avgPSCLF}, the y-axis, representing the average execution time of PSCLF, is expressed according to the latency of SCL.
For $P=1$, the average execution time of SCLF and SCLF embedding the \gls{llrm} \cite{LLRM} is shown.
\gls{llrm} is a mechanism that reduces the average execution time of SCLF without degrading the error-correction performance.
For the \gls{llrm} results, the restart-location choices are made according to the method} presented in \cite{LLRM} and the number of restart locations is constrained to 4, maximizing the average execution-time reduction.

For $\omega=1$ and $\Tmax=20$, \autoref{fig:exec_psclf20} depicts  $\avgSCLF$ and $\avgPSCLF$, as well as the latency of SCL with $L=2$.
For $\text{FER}\simeq1$, the average execution time of PSCLF is not equal to its latency, which is $20\times\latSCL$ \eqref{eq:latPSCLF}.
This is explained by early termination and  CRC collisions.
The average execution-time reduction in comparison to SCLF is given next.
At $\text{FER}=10^{-2}$, the reduction is of $16$\% and $23$\% for $P=2$ and $P=4$, respectively.

\autoref{fig:exec_psclf50} and \autoref{fig:exec_psclf} depict $\avgSCLF$ and $\avgPSCLF$ for $(\omega,\Tmax)=(2,50)$ and  $(\omega,\Tmax)=(3,300)$.
For $\omega=2$, the average execution time reduction with respect to SCLF is $14$\% and $25$\% for $P=2$ and $P=4$ at $\text{FER}=10^{-2}$.
For $\omega=3$, the reduction is greater, being of $44$\% and $77$\%.

For all decoding orders $\omega$, the average execution time of SCLF embedding the \gls{llrm} with $4$ restart locations approaches the average execution time of the proposed algorithm with $P=2$.
At very high \gls{fer}, the proposed algorithm exhibits the best average execution time since the algorithm is able to early-terminate, which is not the case for \gls{sclf} \eqref{eq:ET_P1}.
\begin{figure}
    \centering
    \resizebox{0.99\columnwidth}{!}{\usetikzlibrary{spy}
\begin{tikzpicture}[spy using outlines={circle, magnification=2, connect spies}]
  \pgfplotsset{
    label style = {font=\fontsize{10pt}{8.2}\selectfont},
    tick label style = {font=\fontsize{8pt}{8.2}\selectfont}
  }

   \begin{semilogyaxis}[%
    width=\columnwidth,
    height=0.75\columnwidth,
    xmin=1e-04, xmax=1e-00,
    xlabel={Frame-error rate},
    xlabel style={yshift=0.4em},
    ymin=2000, ymax=84000,
    x dir=reverse,
    xmode=log,
    ytick = {2624,5248,10496,20992, 41986,83972},
    yticklabels={$\latSCL$,2$\latSCL$,4$\latSCL$,8$\latSCL$,16$\latSCL$,32$\latSCL$},
    ylabel style={yshift=0.4em},
    ylabel={Avg. Exec. Time},
    xlabel style={yshift=0.2em},
    yminorticks, xmajorgrids,
    ymajorgrids, yminorgrids,
    legend style={at={(0.99,1.0)},anchor=north east},
    legend style={legend columns=2, font=\scriptsize, column sep=0mm, row sep=-1mm}, 
    legend cell align={left},
    mark size=1.8pt, mark options=solid,
    ]

  \addplot [dashed, color=black, line width=0.8pt, mark size=2.1pt]table[row sep=crcr]{
  3.2e-04   2624\\      
  1e-00   2624\\
  };  
  \addlegendentry{SCL-2}

           \addplot[color=matlab2,mark=o, line width=0.8pt, mark size=2.1pt]
   table[x=FER,y=cc]{PSCLF1_L2_1024_544_16_16.txt};   
    \addlegendentry{P=2}
\addplot [color=matlab1, mark=x, line width=0.8pt, mark size=2.1pt]
  table[x=FER,y=cc]{PSCLF1_L2_1024_544_8_8_8_8.txt};
  \addlegendentry{$P=4$}   
    \addplot [color=black, line width=0.8pt, mark size=2.1pt]
  table[x=FER,y=ccnogrm]{SCLF1_LLRM_GRM_1024_544.txt};
  \addlegendentry{SCLF }   
  \addplot [color=black,dotted, line width=0.8pt, mark size=2.1pt]
  table[x=FER,y=ccllrm]{SCLF1_LLRM_GRM_1024_544.txt};
  \addlegendentry{SCLF - LLRM}  
  \end{semilogyaxis}


\end{tikzpicture}}
    \caption{Average execution time for $(N,K)=(1024,512)$ with $\omega=1$.}
    \label{fig:exec_psclf20}
\end{figure}
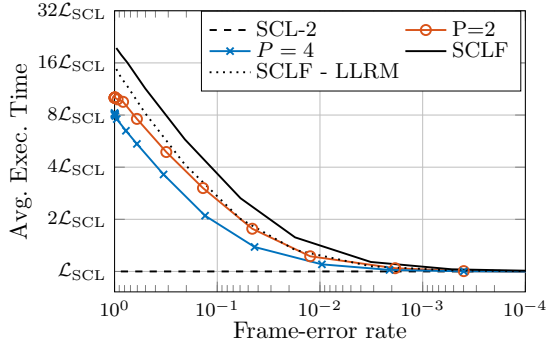
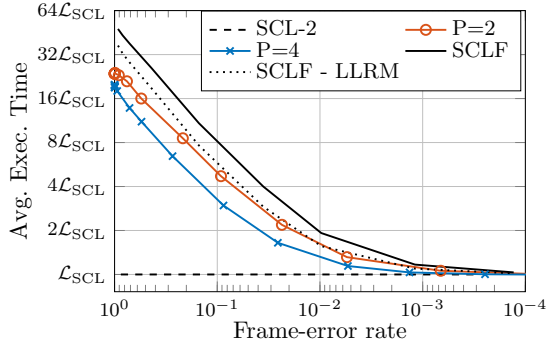
\begin{figure}
    \centering
    \resizebox{0.99\columnwidth}{!}{\usetikzlibrary{spy}
\begin{tikzpicture}[spy using outlines={circle, magnification=2, connect spies}]
  \pgfplotsset{
    label style = {font=\fontsize{10pt}{8.2}\selectfont},
    tick label style = {font=\fontsize{8pt}{8.2}\selectfont}
  }

   \begin{semilogyaxis}[%
    width=\columnwidth,
    height=0.75\columnwidth,
    xmin=1e-04, xmax=1e-00,
    xlabel={Frame-error rate},
    xlabel style={yshift=0.4em},
    ymin=2000, ymax=168000,
    x dir=reverse,
    xmode=log,
    ytick = {2624,5248,10496,20992, 41986,83972,167944},
    yticklabels={$\latSCL$,2$\latSCL$,4$\latSCL$,8$\latSCL$,16$\latSCL$,32$\latSCL$,64$\latSCL$},
    ylabel style={yshift=0.4em},
    ylabel={Avg. Exec. Time},
    xlabel style={yshift=0.2em},
    yminorticks, xmajorgrids,
    ymajorgrids, yminorgrids,
    legend style={at={(0.99,1.0)},anchor=north east},
    legend style={legend columns=2, font=\scriptsize, column sep=0mm, row sep=-1mm}, 
    legend cell align={left},
    mark size=1.8pt, mark options=solid,
    ]

  \addplot [dashed, color=black, line width=0.8pt, mark size=2.1pt]table[row sep=crcr]{
  3.2e-04   2624\\      
  1e-00   2624\\
  };  
  \addlegendentry{SCL-2}

    \addplot[color=matlab2,mark=o, line width=0.8pt, mark size=2.1pt]
   table[x=FER,y=cc]{PSCLF2_L2_1024_544_16_16.txt};   
    \addlegendentry{P=2}
  
  \addplot [color=matlab1,mark=x, line width=0.8pt, mark size=2.1pt]
  table[x=FER,y=cc]{PSCLF2_L2_1024_544_8_8_8_8_CK.txt};
  \addlegendentry{P=4}   
    \addplot [color=black, line width=0.8pt, mark size=2.1pt]
  table[x=FER,y=ccnogrm]{SCLF2_LLRM_GRM_1024_544.txt};
  \addlegendentry{SCLF}   
  \addplot [color=black,dotted, line width=0.8pt, mark size=2.1pt]
  table[x=FER,y=ccllrm]{SCLF2_LLRM_GRM_1024_544.txt};
  \addlegendentry{SCLF - LLRM}   
  \end{semilogyaxis}


\end{tikzpicture}}
    \caption{Average execution time for $(N,K)=(1024,512)$ with $\omega=2$.}
    \label{fig:exec_psclf50}
\end{figure}

\subsection{Check and Keep vs Check and Remove PSCLF}
In \cite{PPC}, the \gls{psclf} decoding algorithm embedded the \gls{ck} method \cite{5G_DCA_WCNC}, which is the method that does not require any change at the start of the partition.
In \gls{ck}, all path metrics are unchanged at the start of the next partition.
In \autoref{subsec:psclf_description_crck}, we propose to also investigate the \gls{cr} method \cite{5G_DCA_WCNC} for \gls{psclf}.
Namely, at the start of the next partition, only the paths checking the \gls{crc} pursue the decoding.
This is performed in \gls{psclf} by simply adding a large constant to the path metrics \eqref{eq:pm_scl_CR} corresponding to paths which do not pass the \gls{crc} at the end of the previous partition. 

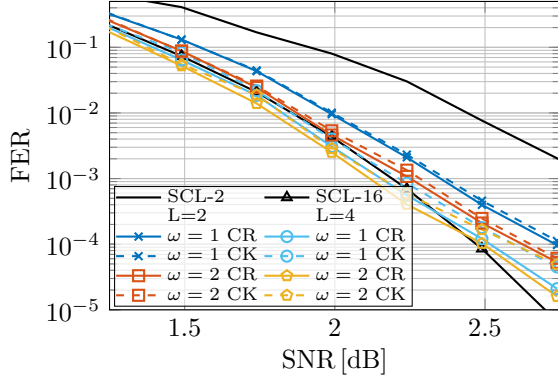
\begin{figure}[t!]
    \centering
    \vspace{2pt}
    \resizebox{.99\columnwidth}{!}{\begin{tikzpicture}
  \pgfplotsset{
    label style = {font=\fontsize{10pt}{8.2}\selectfont},
    tick label style = {font=\fontsize{10pt}{8.2}\selectfont}
  }

   \begin{semilogyaxis}[
    width=\columnwidth,
    height=0.75\columnwidth,
    xmin=1.25, xmax=2.75,
    xlabel={$\text{SNR}\,[\mathrm{dB}]$},
    xlabel style={yshift=0.4em},
    ymin=1e-5, ymax=0.5,
    ylabel style={yshift=-0.1em},
    ylabel={FER},
    yminorticks, xmajorgrids,
    ymajorgrids, yminorgrids,
    legend style={at={(0.01,0.01)},anchor=south west},
    legend style={legend columns=2, font=\footnotesize, row sep=-1mm},
    legend style={fill=white, fill opacity=0.7, draw opacity=1,text opacity=1}, 
    legend cell align={left},
    legend style={inner xsep=0.2pt, inner ysep=-1pt}, 
    mark size=1.6pt, mark options=solid,
    ]   
           \addplot[color=black,line width=0.8pt, mark size=2.1pt]
    table[x=SNR,y=SCL2]{SCL_1024_512.tex};
    \addlegendentry{SCL-2}
    \addplot[color=black, mark=triangle,line width=0.8pt, mark size=2.1pt]
    table[x=SNR,y=SCL16]{SCL_1024_512.tex};
    \addlegendentry{SCL-16}
    \addlegendimage{empty legend}
    \addlegendentry{L=2}
    \addlegendimage{empty legend}
    \addlegendentry{L=4}

\addplot[color=matlab1,mark=x, line width=0.8pt, mark size=2.1pt]
   table[x=SNR,y=FER]{PSCLF1_L2_1024_544_8_8_8_8.txt}; 
    \addlegendentry{$\omega=1$ CR}

    \addplot[color=matlab6,mark=o, line width=0.8pt, mark size=2.1pt]
   table[x=SNR,y=FER]{PSCLF1_L4_1024_544_8_8_8_8_CR.txt}; 
    \addlegendentry{$\omega=1$ CR }
   \addplot[color=matlab1,dashed,mark=x, line width=0.8pt, mark size=2.1pt]
   table[x=SNR,y=FER]{PSCLF1_L2_1024_544_8_8_8_8_CK.txt};   
    \addlegendentry{$\omega=1$ CK}
   \addplot[color=matlab6,dashed,mark=o, line width=0.8pt, mark size=2.1pt]
   table[x=SNR,y=FER]{PSCLF1_L4_1024_544_8_8_8_8_CK.txt};   
    \addlegendentry{$\omega=1$ CK }

           \addplot[color=matlab2,mark=square, line width=0.8pt, mark size=2.1pt]
   table[x=SNR,y=FER]{PSCLF2_L2_1024_544_8_8_8_8.txt};   
    \addlegendentry{$\omega=2$ CR}
    \addplot[color=matlab3,mark=pentagon, line width=0.8pt, mark size=2.1pt]
   table[x=SNR,y=FER]{PSCLF2_L4_1024_544_8_8_8_8_CR.txt}; 
    \addlegendentry{$\omega=2$ CR }
       \addplot[color=matlab2,dashed,mark=square, line width=0.8pt, mark size=2.1pt]
   table[x=SNR,y=FER]{PSCLF2_L2_1024_544_8_8_8_8_CK.txt};   
    \addlegendentry{$\omega=2$ CK}

   \addplot[color=matlab3,dashed,mark=pentagon, line width=0.8pt, mark size=2.1pt]
   table[x=SNR,y=FER]{PSCLF2_L4_1024_544_8_8_8_8_CK.txt};   
    \addlegendentry{$\omega=2$ CK }

  \end{semilogyaxis}

\end{tikzpicture}}
    \caption{Error-correction performance with CR and CK methods with $(N,K)=(1024,512)$.}
    \label{fig:FER_CR_CK}
\end{figure}

The \gls{cr} method is expected to perform better than the \gls{ck} method.
The list size was limited to $L=2$ to reduce the chance of CRC collisions, we also provide in this section performance for $L=4$.
The code parameters are $(N,K)=(1024,512)$, and $P=4$ partitions and $\omega=\{1,2\}$ are used.
Error-correction performance is depicted in \autoref{fig:FER_CR_CK}.
For $L=2$, the performance is slightly improved by using the \gls{cr} method. For $\omega=2$, the gain is around $0.04$\,dB at $\text{FER}=10^{-3}$.
For $L=4$, the gain is greater, especially at low \gls{fer}. At $\text{FER}=10^{-4}$, the gain is of approximately $0.1$\,dB with respect to the \gls{ck} method, for both $\omega=\{1,2\}$.

\section{Conclusion}
In this paper,  the \acrfull{psclf} decoding algorithm is further investigated.
In \gls{psclf}, the code is divided into partitions and each partition is decoded with the SCLF decoder using the dynamic flip metric.
The memory model, taking into account the block to handle the partitioning of the polar codes, shows that the memory overhead is limited with respect to the plain \gls{sclf}.
The probability of \gls{crc} collisions and early-termination under \gls{psclf} is characterized for all decoding parameters. 
Numerical results show that the proposed \gls{psclf} algorithm gains up to 0.1\,dB with respect to \gls{sclf} in terms of \gls{fer}.
This gain is obtained with a design of partitions shown to be valid for three different code rates.
The \gls{crc} structure is shown to have a limited impact, as the chance of \gls{crc} collisions increases with the decoding order $\omega$.
A low-complexity method to restart the decoding of a partition permits a gain up to 0.1\,dB with respect to the restart method shown in \cite{PPC}.
The average execution time of \gls{psclf} was estimated to be $77$ \%  lower than that of \gls{sclf} at a \gls{fer} of $0.01$.
Finally, the average execution time of the proposed decoding algorithm matches the latency of \gls{scl} at low \gls{fer}, while achieving an error-correction performance close to that of \gls{scl}-16 and \gls{scl}-64, for $\omega=2$ and $\omega=3$, respectively.

\section{Acknowledgments}
The authors declare no conflicts of interest.

The original contributions presented in the study are included in the article, further inquiries can be directed to the corresponding author/s.
\bibliography{IEEEabrv,ConfAbrv,references}
\end{document}